%% file: paper-rev-2.tex
\documentclass[prd,twocolumn,showpacs,linenumbers,groupedaddress,superscriptaddress,amsmath,amssymb]{revtex4}

\usepackage{graphicx}
\usepackage{dcolumn}
\usepackage{bm}
\usepackage{amssymb}
\usepackage{enumerate}
\usepackage{lineno}
\usepackage{multirow} 

\newcommand\ee{e^+e^-}

\newcommand\ainv{A'\to invisible}

\newcommand\g{\gamma}
\newcommand\ma{m_{A'}}

\newcommand\na{{n}_{A'}}
\newcommand\Na{{N}_{A'}}
\newcommand\ea{e^- Z \to e^- Z A';~ A' \to \chi \overline{\chi}}
\newcommand\emu{e^- Z \to e^- Z \g; \g \to \mu^+ \mu^-}
\newcommand\mm{\mu^+ \mu^-}

\def\address{\@ifstar{\address@star}%
  {\@ifnextchar[{\address@optarg}{\address@noptarg}}}

\begin{document}

\title{Search for  vector mediator of  Dark Matter production  in invisible decay mode
}
\input author_list.tex

\date{\today}


\begin{abstract}
  A  search  is performed for  a new sub-GeV vector  boson ($A'$)  mediated  production of  Dark Matter ($\chi$)  in the fixed-target 
  experiment, NA64, at the CERN SPS.  The $A'$, called dark photon, can be  generated  in  the    reaction $ e^- Z \to e^- Z A'$   
  of 100 GeV electrons dumped against an active target followed  by its prompt invisible decay $A' \to \chi \overline{\chi}$. 
   The experimental signature of this process  would be an event with an isolated electron and large missing energy in the detector.    From the analysis of the data sample collected in  2016 corresponding to  $4.3\times10^{10}$  electrons on target no evidence of such a process  has been found. 
New stringent constraints on the $A'$ mixing strength with photons,   $10^{-5}\lesssim \epsilon \lesssim 10^{-2}$,  for the $A'$ mass range
 $ \ma \lesssim 1$ GeV are derived.  For  models considering scalar and fermionic thermal Dark Matter interacting with the visible sector  through the vector portal the 90\%  C.L.  limits  $10^{-11}\lesssim y \lesssim 10^{-6}$  on the dark-matter parameter  $y = \epsilon^2 \alpha_D (\frac{m_\chi}{m_{A'}})^4 $ 
are  obtained   for  the dark coupling constant $\alpha_D = 0.5$  and dark-matter masses $0.001 \lesssim m_\chi \lesssim 0.5 $ GeV. 
 The lower limits  $\alpha_D \gtrsim 10^{-3} $ for pseudo-Dirac Dark Matter in the mass region 
 $ m_\chi \lesssim  0.05 $ GeV are  more stringent than the corresponding bounds from beam dump experiments.
The  results are obtained by  using  exact tree level  calculations of the  $A'$ production cross-sections, which turn out to be significantly smaller 
 compared to the one obtained in the  Weizs\"{a}cker-Williams  approximation for the mass region $\ma \gtrsim 0.1$ GeV.
\end{abstract}

\pacs{14.80.-j, 12.60.-i, 13.20.Cz, 13.35.Hb}
\maketitle


\section{Introduction}
\label{sec:intro}
Despite the intensive experimental searches  dark matter (DM) still is a great
puzzle. The difficulty so far is that DM can be probed only through its gravitational interaction with visible  matter.
An  exciting possibilities  is that in addition to gravity,  a new force between the  dark and visible matter transmitted by a new vector  boson,  $A'$ , called dark photon, might  exist ~\cite{Fayet,prv,ArkaniHamed:2008qn,jr}. 
The $A'$ can  have a mass $m_{A'}\lesssim 1$ GeV,  and couple to 
the standard model (SM)  via kinetic mixing  with the ordinary photon,  
described by the  term  $\frac{\epsilon}{2}F'_{\mu\nu}F^{\mu\nu}$ and  parameterized by the mixing strength  $\epsilon$. The Lagrangian of the SM is extended by  the dark sector in the following way:
\begin{eqnarray}
\mathcal{L} = \mathcal{L}_{SM}  -\frac14 F'_{\mu\nu}F'^{\mu\nu}+
\frac{\epsilon}{2}F'_{\mu\nu}F^{\mu\nu}+  \frac{m_{A'}^2}{2}A'_\mu A'^\mu \nonumber \\
+ i \bar{\chi}\gamma^\mu \partial_\mu \chi -m_\chi \bar{\chi} \chi -e_D \bar{\chi}\gamma^\mu A'_\mu \chi, 
\label{DarkSectorLagrangian}
\end{eqnarray}
where the  massive vector field $A'_\mu$ is associated with the  spontaneously broken 
$U_D(1)$ gauge group, 
$F'_{\mu\nu} = \partial_\mu A'_\nu-\partial_\nu A'_\mu$, $e_D$ is the coupling constant of the $U(1)_D$ gauge interactions, and $m_{A'},~m_\chi$  are the masses of the dark photon and DM particles, respectively.
 Here, we consider as an
example the Dirac spinor field, $\chi$, which is treated as Dark 
Matter fermions  coupled to 
$A'_{\mu}$ by  the dark portal 
  coupling constant  $e_D$.   The mixing term  of ~\eqref{DarkSectorLagrangian}
 results in the interaction:
\begin{equation}
\mathcal{L}_{int}= \epsilon e A'_{\mu} J^\mu_{em}
\end{equation}
of dark photons with the electromagnetic current $J^\mu_{em}$ with a strength $\epsilon e$, 
where $e$ is the electromagnetic coupling and $\epsilon \ll 1$ \cite{Okun:1982xi,Galison:1983pa,Holdom:1985ag}. Such small values of $\epsilon$ 
can be obtained naturally in GUT from loop effects  of particles charged under both the dark and SM  $U(1)$ interactions with  a typical 1-loop  value  $\epsilon = e e_D/16\pi^2 \simeq 10^{-2}-10^{-4}$ \cite{Holdom:1985ag}, while  2-loop contributions result in the range 
$10^{-3}-10^{-5}$. An additional hint for the existence of $A'$ is suggested  by the 3.6 $\sigma$ deviation from the SM prediction of the muon anomalous magnetic moment $g_\mu-2$ \cite{g-2},  which can be explained by a sub-GeV  $A'$ with coupling $\epsilon\simeq 10^{-3}$ \cite{gk,Fayet:2007ua,Pospelov:2008zw}, 
 as well as by  hints of astrophysical signals of DM \cite{ArkaniHamed:2008qn}. 
This has motivated a worldwide experimental and theoretical effort  towards  dark forces and other portals between the visible and dark sectors, see Refs. \cite{jr,Essig:2013lka,report1,report2, pdg} for a review.

\par Since there are no firm predictions for  the $A'$, its 
experimental searches  have been performed   over a wide range of $A'$ masses and decay modes.
If the $A'$ is the lightest state in the dark sector,  then it would decay mainly visibly, i.e., typically to SM leptons $l$ (or hadrons) with the rate given by
\begin{equation}
\Gamma (A'\rightarrow  l^- l^+) = \frac{\alpha  \epsilon^2}{3} m_{A'}\Bigl(1+
\frac{2m_l^2}{m_{A'}^2}\Bigr)\sqrt{1-\frac{4m_l^2}{m_{A'}^2}}, 
\end{equation}
 which can be used to detect it. Here, $\alpha = e^2/4\pi$  and $m_l$ is the lepton mass. 
 Such dark photons in the  mass region below a few $\rm GeV$ has been mainly  searched for in  beam dump, fixed target, collider and rare meson decay  experiments, which already put stringent  limits  on  the  mixing $\epsilon^2 \lesssim 10^{-7}$ of such dark photons excluding,  in particular,  
  the parameter region favored by the $g_\mu-2$ anomaly \cite{pdg}-\cite{kloe3}.\\ 

 However, in the presence of light dark states $\chi$,  in particular, DM with the masses $m_\chi<\ma/2$, the $A'$  would predominantly  decay invisibly  into those particles provided that  coupling $g_D >\epsilon e$. 
The decay rate of  $A'\rightarrow \bar{\chi} \chi$ in this case is  given by
 \begin{equation}
\Gamma (A'\rightarrow \bar{\chi}\chi) = \frac{\alpha_D}{3} m_{A'}\Bigl(1+
\frac{2m_\chi^2}{m_{A'}^2}\Bigr)\sqrt{1-\frac{4m_\chi^2}{m_{A'}^2}}.
\end{equation}
Various dark sector models motivate  sub-GeV scalar and Majorana or  pseudo-Dirac fermion DM coupled to dark photons 
\cite{report2,deNiverville:2011it,pdg,Izaguirre:2014bca,Iza2015,Iza2017, kuflic1, kuflic2,kuflic3}.  To 
interpret the observed  abundance of thermal relic density, 
the requirement of the  thermal freeze-out of DM annihilation into visible matter 
through $\g-A'$ kinetic mixing allows one to derive a relation among the parameters  
\begin{equation}
\alpha_D \simeq 0.02 f \Bigl( \frac{10^{-3}}{\epsilon}\Bigr)^2\Bigl( \frac{\ma}{100~ MeV}\Bigr)^4
\Bigl( \frac{10~MeV}{m_\chi}\Bigr)^2
\label{alphad}
\end{equation}
where $\alpha_D = e_D^2/4\pi$, $f \lesssim 10$ for a 
scalar \cite{deNiverville:2011it}, and $f\lesssim 1$ for a  fermion 
\cite{Izaguirre:2014bca}. This 
 prediction combined with the fact  that  the intrinsic scale of the dark sector could be  smaller than, or comparable to, that of the visible sector, provide  an important target for the ($\epsilon, ~\ma$) parameter space which can be probed at energies attainable at the CERN SPS. 
Models introducing such invisible  $A'$ also offered possibilities to explain the $g_\mu-2$ and various other anomalies \cite{Lee:2014tba} and are subject to different experimental   constraints  \cite{ei,Diamond:2013oda,hd,Essig:2013vha}.  The severe limits on invisible   decays of  sub-GeV $A'$s have been obtained from the results of beam dump experiments LSND~ 
\cite{deNiverville:2011it,Batell:2009di} and E137 \cite{e137th},  under assumptions of certain values of the coupling strength,  $\alpha_D$,  and masses of the DM decay particles.  Recently,  NA64 \cite{na64prl} and BaBar \cite{babarg-2} experiments set new direct limits, $\epsilon^2 \lesssim10^{-6}$ for $m_{A'} \lesssim 100$  MeV and $\epsilon^2 \lesssim10^{-6}$ for $m_{A'} \lesssim 1$ GeV, respectively,   which rule out the $A'$ parameter space 
explaining the muon g-2 anomaly,  leaving,  however,  a  significant  area that is still unexplored. 

In the following we assume  that the Dark Matter invisible decay mode is dominant, i.e. $\Gamma (A'\rightarrow \bar{\chi}\chi)/\Gamma_{tot} \simeq 1$, and 
 that the $A'$ leptonic decay channel is suppressed, $\Gamma (A'\rightarrow \bar{\chi}\chi) \gg \Gamma (A'\rightarrow  e^- e^+)$.
  If such  $A'$ exists, many crucial questions about its mass scale, coupling constants, decay modes, etc. arise. One possible way to answer these questions, is  to search  for the invisible $A'$ in accelerator experiments. The $A'$s  could be 
produced in a high intensity beam dump  experiment and generate a flux of DM particles  through their decays,  which can be detected through the  scattering off electrons in  the detector target \cite{deNiverville:2011it,Izaguirre:2014bca,Diamond:2013oda,Batell:2009di,Dharmapalan:2012xp,Gninenko:2012prd,Gninenko:2012plb}.  In this case, the 
signal event rate in the detector  scales as $\epsilon^2 y \propto \epsilon^4 \alpha_D$, where  the parameter $y$ is  defined as 
\begin{equation}
y = \epsilon^2 \alpha_D \Bigl(\frac{m_\chi}{m_{A'}}\Bigr)^4, 
\label{y}
\end{equation}
which was recently 
constrained to $10^{-9}\lesssim y \lesssim 10^{-8}$, for $\alpha_D = 0.5$ 
and for dark-matter masses of $0.01 < m_\chi < 0.3$ GeV  by the MiniBooNE  experiment~ \cite{minib2017}.

Another approach, considered in this work and proposed in Refs.~\cite{Gninenko:2013rka,Andreas:2013lya}, 
 is based on the detection of the large missing energy, carried away by the energetic $A'$
 produced in the interactions of high-energy electrons in the active beam dump target, see also \cite{Izaguirre:2014bca}.   The advantage of this type of experiment is that its sensitivity is proportional to the mixing strength squared, $\epsilon^2$,  associated with the $A'$ production  in the primary reaction and its subsequent prompt invisible decay, while in the former case it is proportional to $\epsilon^4\alpha_D$, with $\epsilon^2$ associated with  the $A'$ production in the beam dump and  $\epsilon^2 \alpha_D$ coming from the $\chi$ particle interactions in the detector.
 \par In this work  we report new results on the search for  the $A'$ and light DM in the fixed-target  experiment NA64 at the CERN SPS.
 The experimental signature of events from the $\ainv$ decays is clean and they can be selected with small background
due to the excellent capability of NA64 for the  precise identification and measurements of the 
initial electron state.
\par The rest of the paper is organised as follows. 
Section \ref{sec:method} outlines the method of search and theoretical setup for the $A'$ production in an 
electron- nuclei scattering,  and the signal simulation. 
Here,  we mainly focus   on the  experimental signature  of the $\ainv$ decays and $A'$ production rate. 
 We also attempt to provide an estimate of the experimental  uncertainties associated with the $A'$  cross section  calculation required for the sensitivity estimate.   We revisit here the calculations of 
Refs.\cite{Gninenko:2013rka,Andreas:2013lya,gkkk} and  
clarify the apparent disagreements in  the numerical factors in the cross section  of 
$A'$ production   in the Weizs\"{a}cker-Williams  framework and exact computations at tree level.
We also discuss additional experimental inputs that would be useful to improve the reliability of the calculated sensitivity of the NA64 experiment. 
The H4 beam line and experimental set-up is presented in Sec. \ref{sec:detector}, followed by a description of the event reconstruction and analysis  in Sec. \ref{sec:analysis}. 
 The results  on  the benchmark process  of  dimuon  production are presented in  Sec.\ref{sec:dimuon}. In Sec. \ref{sec:eff} and  \ref{sec:bckg} the signal efficiency 
and background  sources are discussed. The final results on the searches for invisible decays 
of dark photons and light thermal DM are  reported  in 
 Sec. \ref{sec:results} and  \ref{sec:ltdm}, respectively.  We present our conclusions  in Sec. \ref{sec:conclusion}.

\section{Method of search and the $A'$ production}
\label{sec:method}
As seen from the Lagrangian \eqref{DarkSectorLagrangian}, any source of  photons   will 
produce all kinematically possible  massive $A'$ states according to the appropriate mixing strength.
If the  coupling strength $\alpha_D $ and $A'$ masses are as discussed above, 
the $A'$ will  decay predominantly invisibly.  

\begin{figure}
\includegraphics[width=0.5\textwidth]{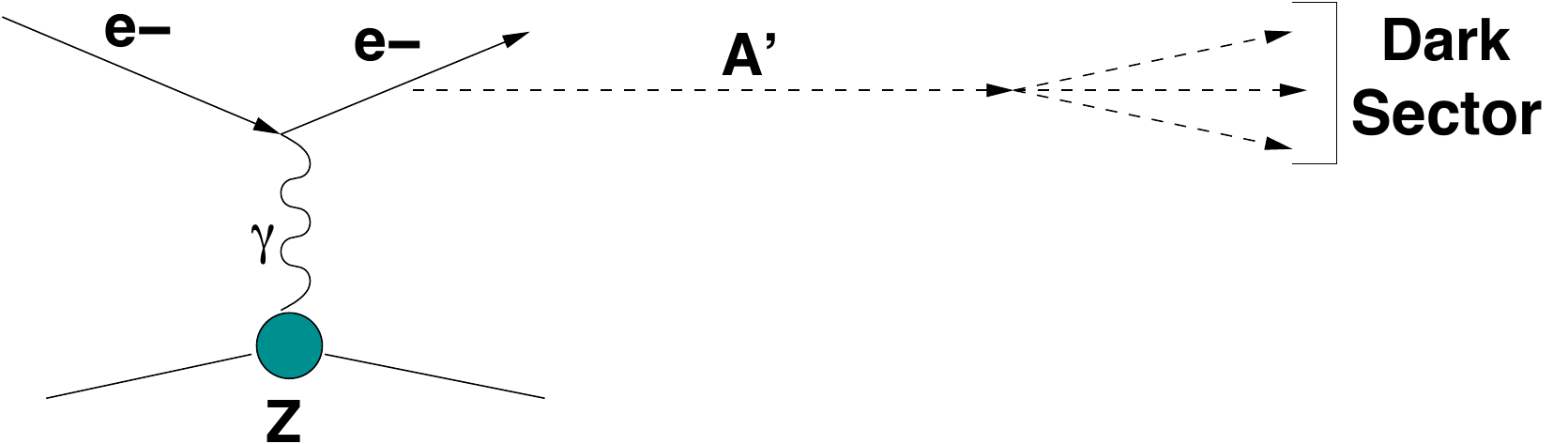}
\caption {Diagram contributing to the $A'$ production in the  reaction  
 $e^- Z \rightarrow e^-  Z A',  A' \rightarrow dark~sector$. The  produced $A'$  decays invisibly into dark sector particles.
 \label{diagr}}
\end{figure}

The method of the search for the $\ainv$ decay is as follows \cite{Gninenko:2013rka,Andreas:2013lya}. If the $A'$ exists it could be produced via the kinetic mixing with bremsstrahlung photons  in the reaction
 of high-energy electrons absorbed in  an active beam dump (target)   followed by the prompt $\ainv$  decay into DM particles  in  a hermetic detector:
\begin{equation}
\ea
\label{e-a}, 
\end{equation} 
see Fig.~\ref{diagr}.
  A fraction $f$ of the  primary beam energy  $E_{A'} = f E_0$ is carried away by $\chi$ particles,  which penetrate the target and  
detector without interactions  resulting in  zero-energy deposition. The remaining part of the beam energy $E_e=(1-f)E_0$ is deposited in the target  by the  scattered electron. 
The  occurrence of the $A'$  production  via  the reaction \eqref{e-a} would appear as an excess of events with a signature of  a single isolated electromagnetic (e-m) shower in the dump  with energy $E_e$ accompanied by a  missing energy $E_{miss}=E_{A'}= E_0 - E_e$  above those expected from backgrounds. Here we assume that in order to give a missing energy signature the $\chi$s have to traverse the detector without  decaying visibly. No other assumptions are made on the nature of the $\ainv$ decay . 
In  previous work \cite{na64prl,gkkk},  the   differential cross-section  $A'$-production  from  reaction  \eqref{diagr} was calculated with 
 the  Weizs\"{a}cker-Williams (WW) approximation, see  \cite{jb,Tsai:1986tx}.
The cross-sections were implemented a Geant4 \cite{Agostinelli:2002hh,geant} based simulations,  and  the total number  $\na$ of the  produced $A'$ per single electron on target (EOT),  depends in particular on $\epsilon,~m_{A'},~E_0$  and was calculated  as 
\begin{equation}
\na(\epsilon,~m_{A'},~E_0)   =  \frac{\rho N_{A}}{A_{{\rm Pb}}}   \sum_i   n(E_0,E_e,s) \sigma^{A'}_{WW}(E_{e}) \Delta s_i
\label{AprYields}
\end{equation} 
where  $ \rho$ is density of Pb target, $N_A$ is the 
Avogadro's number, $A_{{\rm Pb}}$ is the  
Pb atomic mass, $n(E_0,E_e,s)$ is  the number of $e^\pm$ with the energy $E_e$ in the e-m shower at the depth $s$ (in radiation lengths) within the target of total thickness $T$, and $\sigma(E_e)$ is the cross section of the $A'$ production in the kinematically allowed region up to 
$E_{A'}\simeq E_e$ by an electron with the energy $E_e$ in the elementary reaction \eqref{e-a}. 
The  energy distribution $\frac{d n_{A'}}{d E_{A'}}$  of the $A'$s   was  calculated by taking into account the  differential cross-section $\frac{d\sigma(E_e, E_{A'})}{dE_{A'}}$, as  described in Ref.\cite{gkkk}.
\begin{figure}[tbh]
\begin{center}
\vspace{-0.cm}{\includegraphics[width=.5\textwidth]{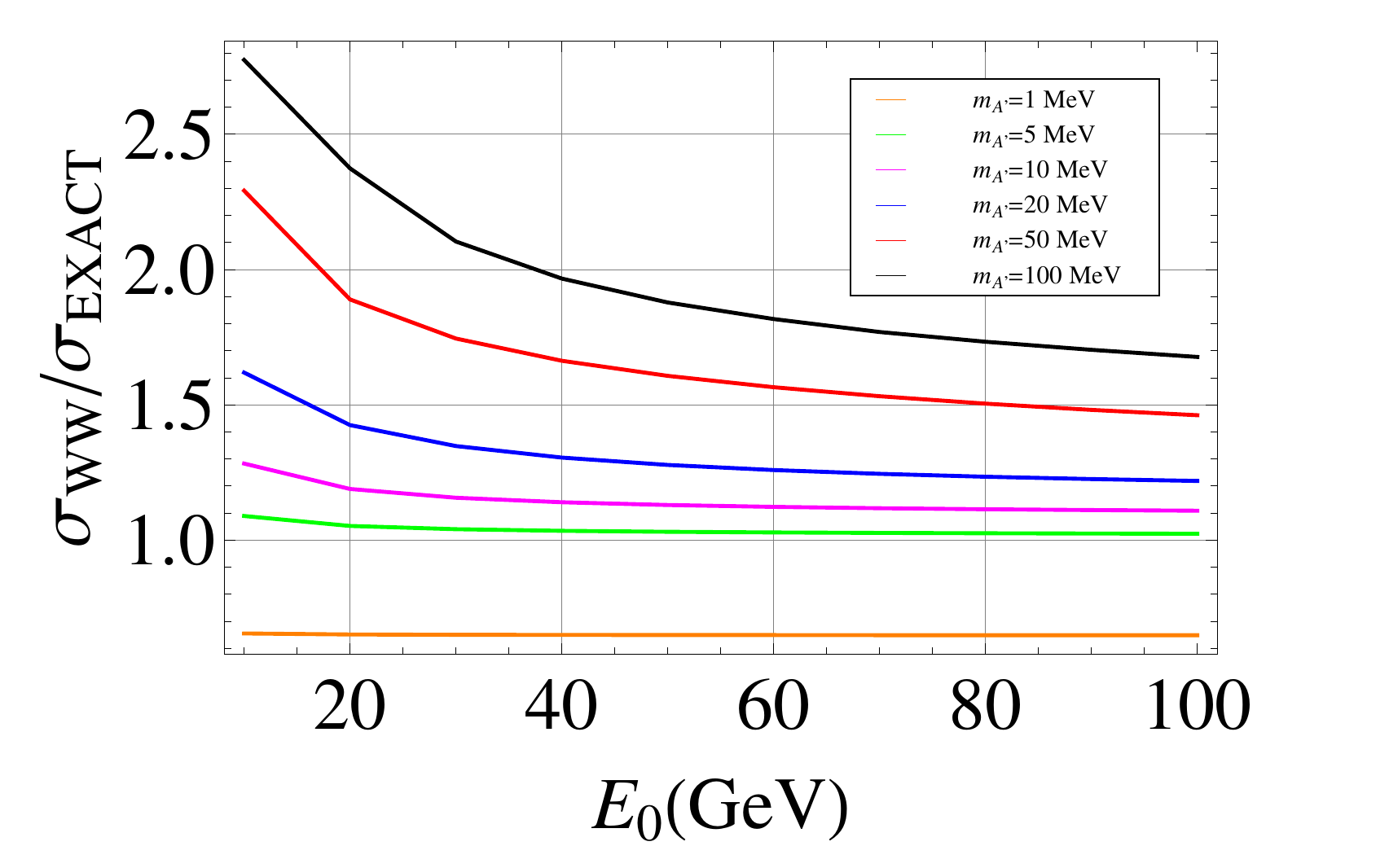}}
\caption {The $k$-factor for  the  $A'$ production  in the reaction  $e^- Z \rightarrow e^-  Z A'$  as a function of  the  electron  energy $E_0$ for different values of 
the $A'$ masses. \label{kfactors}}
\end{center}
\end{figure}
The numerical  summation  in Eq.~(\ref{AprYields}) was 
performed with the detailed simulation of
e-m showers  done by Geant4
over the missing energy spectrum   in the  target, see Fig.~\ref{AspectraForShower}.   According to the simplified 
WW approximation~\cite{jb} the $e^-N$ scattering total rate  
can  be written as 
\begin{equation}
\sigma_{\text{WW}}^{A'} = \frac{4}{3} \frac{\epsilon^2 \alpha^3 \Phi }{m_{A'}^2} \cdot \log \delta^{-1}, \qquad 
\delta = \mbox{max} \left[ \frac{m_e^2}{m_{A'}^2}, \frac{m_{A'}^2}{E_0^2}\right],
\label{TotIWWcrSect1}
\end{equation}
 where
$\Phi$ is the effective flux of photons
\begin{equation}
\Phi   = \int^{t_{max}}_{t_{min}} dt \frac{(t - t_{min})}{t^2}\left[G^{\text{el}}_2(t)+G^{\text{inel}}_2(t)\right]. 
\end{equation}
Here, $t_{min} = m_{A'}^4/(4 E_0^2)$ and $t_{max}=m_{A'}^2$ are 
approximated values of the $A'$ momentum transfer.
For most energies the elastic form-factor $G_{2, el}(t)$ dominates and can be approximated as
\begin{equation}
G_{2,el}(t) = \left(\frac{a^2t}{1 + a^2 t}\right)^2 \left(\frac{1}{1 +t/d}\right)^2  Z^2 \,,
\end{equation}
where $ a = 111~Z^{-1/3}/m_e$ and
$d =$0.164~GeV$^2 A^{-2/3}$.
Note that for heavy atomic nuclei $A$  one has  to take into account  the inelastic nuclear
form factor.  The flux is given by  $\Phi = Z^2 \cdot Log$, where the 
$Log$ value depends weakly on atomic screening,
nuclear size effects  and kinematics 
 \cite{jb,Tsai:1973py}. Numerically 
 $Log \approx (5 - 10) $ for $m_{A'} \leq 500$~MeV.

 It has been recently pointed out, 
that  for a certain kinematic region of the parameters $\ma, E_{A'}$, 
the   $A'$  yield  derived  in the WW framework could  differ 
significantly  from the one obtained with the exact tree-level (ETL) calculations \cite{Liu:2016mqv,Liu:2017htz}.  
Therefore, it is instructive to perform an accurate  
 calculation of the $A'$ cross-section   based on precise phase space integration over the final 
 state  particles  in the reaction  $e^- Z\rightarrow e^-ZA'$.
A reliable  theoretical prediction for the $A'$ yield is essential 
for the proper interpretation of the experimental results  in order to obtain robust exclusion limits in the $A'$ parameter space or 
the possible observation of  the $A'$ signal. 
\par In order to derive  more  accurately  the $A'$ yield,  we have used 
the $A'$ production cross sections of \eqref{e-a} obtained  without  the WW approximation, but  with ETL calculations  of Ref. \cite{kk}. 
These cross sections were cross checked with those calculated  by Liu et al. \cite{Liu:2016mqv,Liu:2017htz} and were found to be in agreement.    The comparison showed that the difference between the two calculations  in  the wide $A'$ mass range  does not exceed  $\simeq10\%$ (see Fig. 1 in Ref.\cite{kk}),  which will be further   used as a systematic  uncertainty  in the calculation  of the $A'$ yield.  This difference is, presumably, due to  the different accuracy of computation programs 
  used for integration over the phase space and, possibly due to the different parameterisation of the form factors  used as an input for the cross section 
  calculations.    
\par
\begin{figure}[tbh!]
\begin{center}
\includegraphics[width=0.5\textwidth,height=0.4\textwidth]{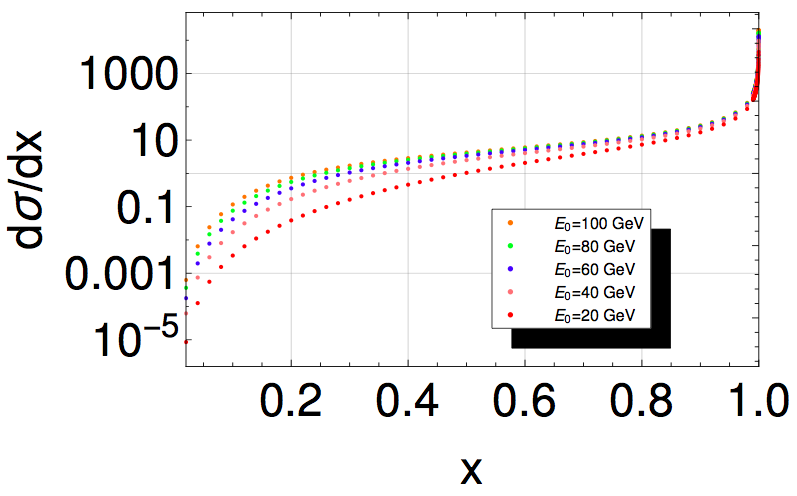}
\caption{ 
The  differential $A'$ spectra from the electron beam interactions in the 
thin ($\ll X_0)$ Pb target calculated for the mass $m_{A'}=100$~MeV as a function of $x=E_{A'}/ E_e$. The spectra are computed  for different electron energies as indicated in the legend. The spectra are normalized to the same number of EOT. 
\label{Aspectrathin}}
\end{center}
\end{figure}

\begin{figure}[tbh]
\begin{center}
\includegraphics[width=0.5\textwidth]{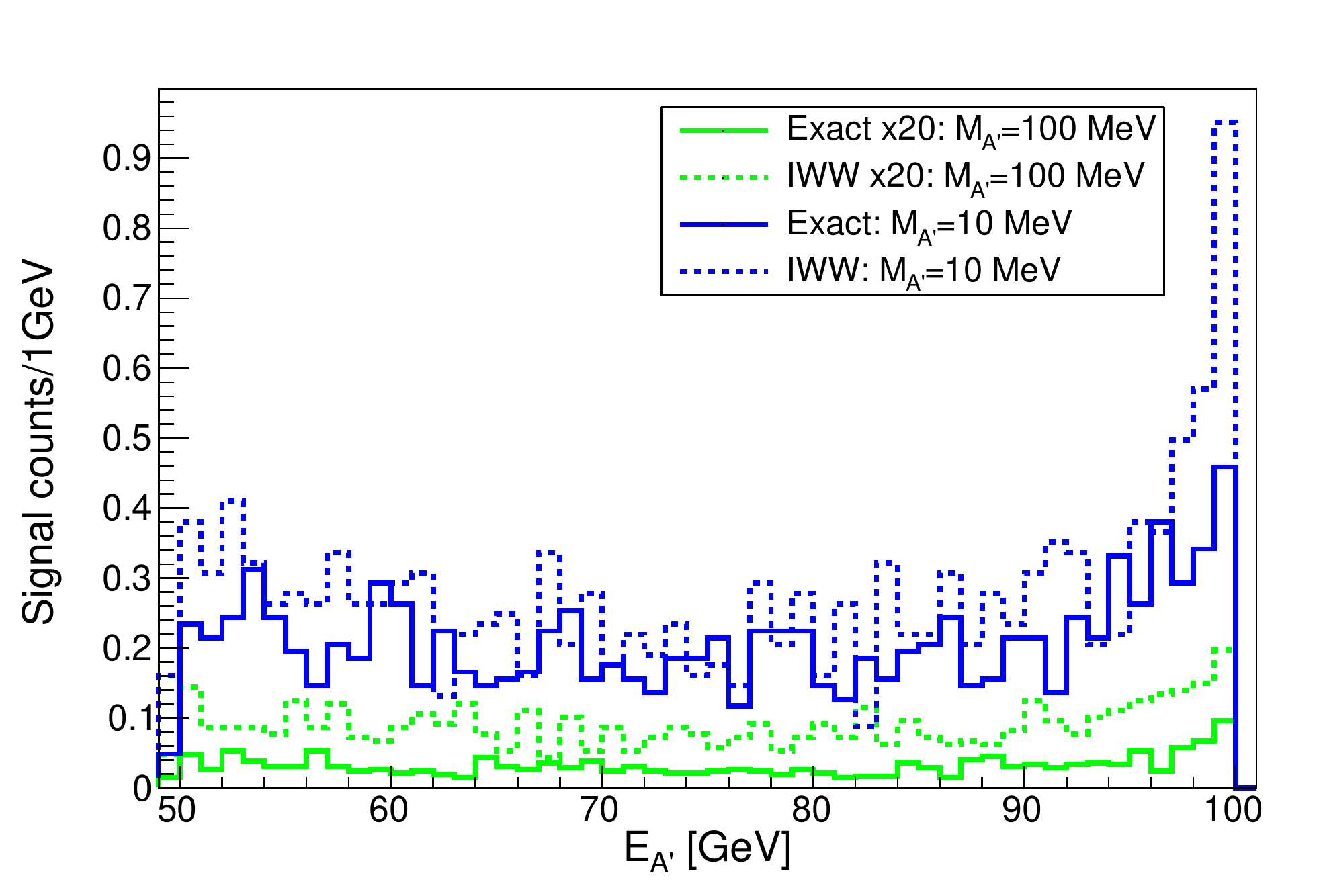}
\caption{ The $A'$ emission spectra from the 100 GeV  electron beam interactions in the thick Pb target ($t\gg X_0$, see Sec.III)) calculated for the masses $m_{A'}=10$ and 100~MeV without  and with  the IWW approximation. The spectra are normalized to  the same number of EOT. 
\label{AspectraForShower}}
\end{center}
\end{figure}

 In order  to implement the ETL cross-section formula into  Geant4~\cite{geant} based NA64  simulation package,   we  introduce  in  Eq. (\ref{AprYields})  a correction $k$-factor defined by the following ratio
\begin{equation}
k(m_{A'},E_0, Z, A) =\frac{\sigma^{A'}_{\text{WW}}}{\sigma^{A'}_{\text{EXACT}}}.
\label{Rdef1}
\end{equation} 
Here, the  cross-section  $\sigma^{A'}_{\text{EXACT}}$ takes into account the 
phase space integration over the final states of the  particles and 
 represents the overall uncertainties in the cross-section (\ref{TotIWWcrSect1}) calculated in simplified WW approach.
The $A'$ yield was calculated by using \eqref{AprYields} with the replacement 
\begin{equation}
\sigma^{A'}_{\text{WW}}  \to k(m_{A'},E_0, Z, A)^{-1} \sigma^{A'}_{\text{WW}}
\label{Rdef2}
\end{equation} 
 We refer to this method as $k$-factor approach throughout the paper.
For heavy target nuclei  $k$ 
depends rather weakly   on $Z$ and $A$, i.e. for tungsten and lead the 
deviation is about $0.5 \%$. In Fig.~\ref{kfactors},  $k$ values are shown as a 
function of the electron beam energy $E_0$ for various $m_{A'}$. 
One can see that  for $m_{A'}=100$ MeV and $E_0=100$ GeV,  the ETL cross sections of the $A'$ 
production are  smaller by a factor $1.7$ than the corresponding WW cross-sections. 
On the other hand, $\sigma_{EXACT}$ exceeds $\sigma_{WW}$ for the  
masses $m_{A'}$ below $5$ MeV, so  that one can slightly improve the limits 
 on  the mixing strength for this mass region. An example of  differential $A'$ spectra from the electron beam interactions in the thin, $\ll X_0$,   
Pb target (here  $X_0$ is the radiation length)
 calculated for the mass $m_{A'}=100$~MeV as a function of $x=E_{A'}/ E_e$  for different electron energies is shown in Fig.~\ref{Aspectrathin}. 

\par Once the $A'$ flux \eqref{AprYields}
was defined, the next step was to simulate  the $A'$ emission spectrum from the target.\ 
The decay electrons and positrons were tracked through  the dump  medium including bremsstrahlung photons, their conversion and 
multiple scattering in the target.\ 
The $A'$  reconstruction efficiency in the target  was computed and convoluted 
with the target details and detector geometrical acceptance  (see Sec. \ref{sec:detector})  based on
the  NA64 Monte Carlo (MC) simulation package  used in our previous search \cite{na64prl}.
The comparison of the  energy distributions of $A'$s emitted  from the thick target ($t\gg X_0$) with the energy $E_{A'}\gtrsim 0.5 E_0$ calculated for masses  $m_{A'}=10$  and 100 MeV with and without WW approximation for the  100 GeV beam energy is shown in Fig.~\ref{AspectraForShower}. The spectra have the similar shape and differ mostly in the overall nomalization factor.  Note that these distributions represent  also the spectra of the  missing energy in the detector. 

 \begin{figure*}[tbh!!]
\includegraphics[width=1.\textwidth,height=.45\textwidth]{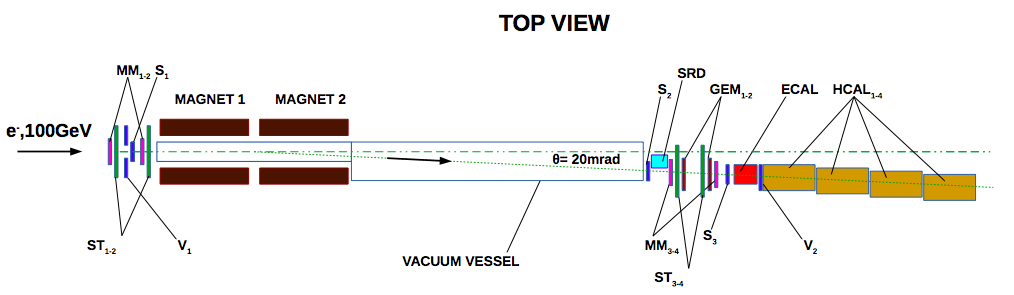}%
\vskip-0.cm{\caption{Schematic illustration of the setup to search for $\ainv$  decays of the bremsstrahlung $A'$s 
produced in the reaction  $eZ\rightarrow eZ A'$ of 100 GeV e$^-$ incident  on the active ECAL target.\label{setup}}}
\end{figure*} 
\section{H4 beam and NA64 detector}
\label{sec:detector}
 The experiment  employs the optimized  100 GeV  electron beam from the H4 beam line at the  North Area (NA) of the CERN SPS described in details in 
Ref.\cite{beam}. The H4 provides an essentially pure $e^-$
beam for fixed-target  experiments.  The  beam was  designed to transport the electrons with the maximal  intensity up to $\simeq 10^7$ per SPS spill of 4.8 s in the  momentum range between 50 and 150 GeV/c that could be produced by the primary proton beam of 400 GeV/c with the intensity up to a few 10$^{12}$ protons on a beryllium target. The main contribution to  the $e^-$ yield  from 
the  target was the production of $\pi^0$ followed by a process $\pi^0 \to \g\g \to \ee$. \
The short-lived $\pi^0$ decays inside the target, and the electrons are  
produced through the conversion of the decay photons in a separate converter 
\cite{h4beam}.  Protons and charged secondaries that  did not  interact in the convertor are separated from the neutrals 
by deflecting  them in a magnetic field   to a  thick absorber. 
The electrons produced in the converter are transported to the NA64 
detector inside an evacuated beam-line tuned to an adjustable  beam momentum.
The hadron contamination in the electron beam was  $\pi/e^- \lesssim 10^{-2}$. The beam  has the transverse size  at the detector  position 
 of the order of a  few cm$^2$ and a halo with intensity  $\lesssim $ a few \%.
\par  The signal event recognition in NA64 must rely on the detection 
of the incoming and outgoing electron only, since the decay product 
 for  the $\ainv$ decay are undetectable.
The  NA64 detector, which is located at about 500 m from the proton target, 
  is schematically shown in Fig.~\ref{setup}.
 The setup utilized 
the beam defining  scintillator (Sc)  counters S1-S3 and veto V1, and  the spectrometer  consisting of two successive  dipole magnets with the integral  magnetic field of $\simeq$7 T$\cdot$m  and low-material-budget tracker. 
The tracker was a set of two upstream Micromegas chambers (MM1,2)  and two downstream MM3,4 and GEM1,2 stations, measuring the beam $e^-$ momenta, $P_e$, with the precision $\delta P_e/P_e \simeq 1\%$ 
\cite{Banerjee:2015eno}.  The Straw Tubes chambers (ST)  were used for calibration purposes.  
The in (out-)coming  electron azimuthal  angle was tuned 
to be within $\theta_{in(out)}\lesssim 1$ mrad with respect to the primary beam axis.  
A small fraction of events with the larger incoming angle in the range  
 $\theta_{in(out)} \simeq1-20$ mrad which was  typically correlated with the smaller  track momentum was rejected by further analysis.  
The magnets also served as an effective filter rejecting the low energy electrons present in  the beam. 
To improve the high energy electron selection  and suppress background from a  possible admixture of low energy electrons,  a tagging system utilizing  the synchrotron radiation (SR) from high energy electrons in the magnetic field  was used,   as shown schematically in Fig.~\ref{setup}. The basic idea was that, since the SR energy emitted by a particle per revolution with a mass $m$ and energy $E_0$ is 
 $<E_{SR}> \propto E_0^3/m^4$,  the low energy electrons and hadrons in the beam could be effectively  rejected by  using the cut on the energy deposited  in the SR detector (SRD)   \cite{Gninenko:2013rka,na64srd}. 
  A 15 m long vacuum vessel was installed  between the magnets and the ECAL to minimize absorption of the SR photons detected immediately at the downstream end of the vessel with a SRD, which was 
an array  of  PbSc sandwich calorimeter with fine longitudinal segmentation.  Compared to the 
previous  measurements \cite{na64prl}, the SRD was also segmented  transversely by   three SRD counters,  each $60\times 80 $mm$^2$
 in lateral size assembled from $80-100~\mu$m Pb and 1 mm Sc plates with wave length shifting  (WLS) fiber read-out.
This allowed to additionally suppress  background from hadrons,  that could knock off electrons from the output
vacuum window of the vessel  producing  a fake $e^-$ SRD tag, 
 by about two  orders of magnitude \cite{na64srd}. 
  The detector was also equipped with an active target, which was a hodoscopic  electromagnetic calorimeter (ECAL) for the measurement of the electron energy deposition,  $E_{ECAL}$, with the accuracy $\delta E_{ECAL}/E_{ECAL} \simeq 0.1/\sqrt{E_{ECAL}[{\rm GeV}]}$ as well as  the $X,~ Y$ coordinates of the incoming electrons by using the transverse e-m shower profile. 
 The ECAL was a  matrix of $6\times 6 $  Shashlik-type counters  
 assembled with  Pb and Sc plates with WLS fiber read-out. Each module was $\simeq 40$ radiation  lengths ($X_0$) and had an initial part $\simeq 4~X_0$ used as a preshower~(PS) detector.  
By requiring the presence of in-time SR signal in  all three  SRD counters,
 and using information of the longitudinal and lateral 
shower development in the ECAL,    the initial level  of the hadron contamination in the  beam $\pi/e^- \lesssim 10^{-2}$ was further suppressed by  more than 4 orders of magnitudes, while keeping 
the electron ID efficiency at the level $\gtrsim 95\%$ \cite{na64srd}.  
The ECAL PMTs were
read-out with sampling ADC (MSADC) electronics which consist of shapers  and
the ADCs themselves \cite{msadc,msadc1}. The shaper stretched the PMT signal to 
$\simeq 100$ ns while the MSADC sampled the signal amplitude every 12.5 ns. As described
below this allowed using an algorithm to 
extract precise timing and amplitude  from  the MSADC information 
in the presence of pileups at high intensity.  The NA64 Data Acquisition system was  adapted from the one used in  the COMPASS experiment at CERN 
\cite{COMPASS_DAQ}.
A  high-efficiency veto counter $V_2$, and a massive, hermetic hadronic calorimeter (HCAL) of $\simeq 30$ nuclear interaction lengths ($\lambda_{int}$) were positioned  just after the ECAL.   The $V_2$  was a plane of scintillation counters   used to
veto  charged secondaries incident on the  HCAL detector from upstream $e^-$ interactions. 
The HCAL which was  an assembly  of four modules HCAL1-HCAL4,  served  as an efficient veto to detect muons or hadronic secondaries produced in the $e^- A$ interactions  in the ECAL target.  Each module was a sandwich of 48 alternating layers of iron and scintillator (Sc) with a thickness of 25 mm and 4 mm,  respectively, with a 
total length  of $\simeq 7\lambda_{int}$, and with a lateral size of $60\times 60$ cm$^2$.  
Each Sc layer consisted  of 3$\times$3  plates with WLS  fiber  readout allowing to assemble the whole HCAL module as a 
matrix of $3\times3$ cells, each of  $20\times 20$ cm$^2$ .
  The number of photoelectrons produced by a minimum ionizing particle (MIP)  crossing the single  module  was in the range $\simeq$ 150-200 photoelectrons.  
The HCAL energy resolution  was $\delta E_{HCAL}/E_{HCAL} \simeq 0.6/\sqrt{E_{HCAL}}[{\rm GeV}]$. 

The single electron events were collected with the hardware trigger  
\begin{equation}
Tr(A') = \Pi S_i \cdot V_1 \cdot PS(>E^{th}_{PS}) \cdot \overline{ECAL}(< E^{th}_{ECAL})
\label{trigger}
\end{equation}
designed to accept events with   in-time hits in  beam-defining counters $S_i$ and clusters in the PS and ECAL with the energy thresholds  $ E^{th}_{PS}\simeq 0.3$ GeV and $E^{th}_{ECAL} \lesssim 80$ GeV, respectively.  
The missing energy  events have the signature 
\begin{equation}
S(A') = Tr(A') \cdot  Track (P_e) \cdot V_2 (< E^{th}_{V}) \cdot HCAL(< E^{th}_{HCAL})
\label{sign}
\end{equation}
with the incoming track momentum $P_e\simeq 100$ GeV, and $V_2$  and HCAL zero-energy  deposition, defined as energy  below the thresholds 
 $ E^{th}_{V_2} \simeq 1$ MIP and  $ E^{th}_{HCAL} \simeq 1$ GeV, respectively.

\section{Data analysis and selection criteria}
\label{sec:analysis}
\par The search for the $\ainv$ decay  described in this paper uses the full data sample
collected  during  July and October runs in 2016 corresponding to  $n_{EOT}=4.3\times 10^{10}$ EOT.
The results reported here are obtained using three sets of data in which
$n_{EOT}=2.3\times 10^{10},~ 1.1\times 10^{10}$ and $0.9\times 10^{10}$ EOT were collected with the beam intensities $\simeq (1.4-2)\times 10^6,~\simeq (3-3.5)\times 10^6$ and 
$\simeq(4.5- 5)\times 10^6$   e$^-$ per spill, respectively.
 Data of these three runs (hereafter called respectively the  run I,II, and III) were processed with  selection criteria similar to the one used in our previous 
paper \cite{na64prl} and finally combined as described in Sec. \ref{sec:results}.
 Compared to the analysis of Ref.\cite{na64prl}, a number of improvements in the event reconstruction, e.g.,  adding the pileup algorithm,  were made in order to increase the 
 reconstruction efficiency. 

 \par The strategy of the analysis was to identify $\ainv$ candidates 
by precise reconstruction of the initial $e^-$ state and an isolated low energy e-m shower in the ECAL   that are accompanied 
by no other activity in the $V_2$ and HCAL detectors.  The measured rate 
of such events  was then supposed to be compared to that expected from known sources.\
\begin{figure}[tbh!!]
\includegraphics[width=.5\textwidth,height=.45\textwidth]{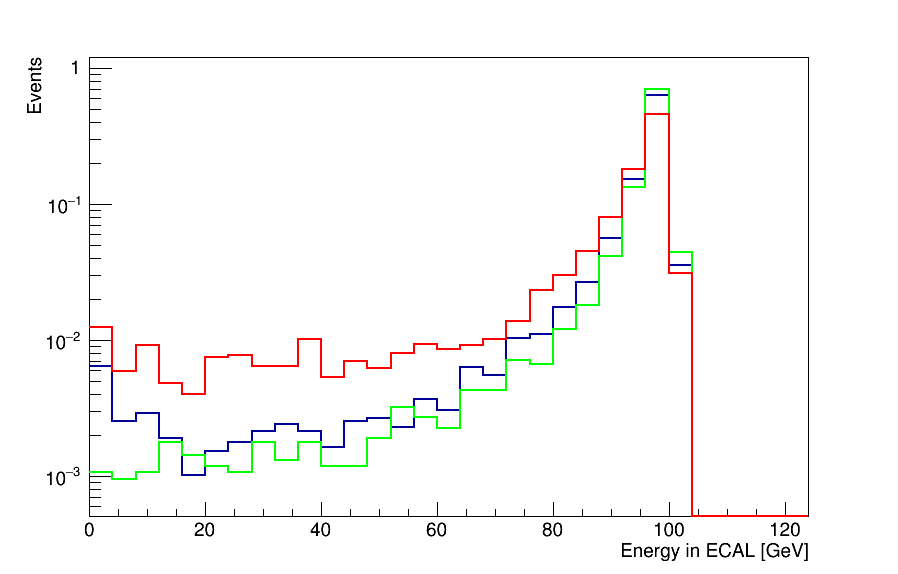}%
\vskip-0.cm{\caption{The MC distributions of energy deposited in the ECAL target from the reaction  $eZ\rightarrow eZ A'$ induced by 100 GeV e$^-$s and 
accompanied by the emission of  the bremsstrahlung $A'$s with the mass 2 (green), 20(blue) and 200 (red) MeV.\label{ecal-spectra}}}.
\end{figure} 
The spectra of $A'$s produced in the ECAL target by primary electrons
were calculated using the approach reported in Ref.\cite{gkkk}.  An example of  the distributions of energy deposited in the target 
calculated for the masses $\ma =2$, 20 and 200 MeV is  shown in  Fig.~\ref{ecal-spectra}. 
A detailed Geant4 based MC simulation  was used to study the detector performance and acceptance losses, 
to simulate background sources, and to select cuts and estimate the reconstruction efficiency.

The candidate events were pre-selected with the criteria chosen to maximize the acceptance for simulated signal events  and to minimize the numbers of events expected from background sources discussed in 
Sec. \ref{sec:bckg}. The following selection criteria were applied:
\begin{itemize}
\item There must be one  and only one incoming particle track  having  a small angle with respect to  the beam axis. This cut rejects low momentum electrons as they  were typically correlated 
with a  large-angle incoming  tracks originating presumably from the upstream $e^-$ interactions. The reconstructed momentum of the particle was required to be $P_e = 100\pm 2 $ GeV. 

\item The  track should be    identified  as an electron  
 with  the SRD detector.
 The energy deposited in each of the three  SRD modules  should be within the SR range emitted by $e^-$s and in time with the trigger. This was the key cut identifying  the pure initial $e^-$ state, with the pion suppression factor  $<10^{-5}$ 
 and electron efficiency $>95\%$ \cite{na64srd}. 

\item The lateral and longitudinal shower shape  in the ECAL should be  consistent with the  one expected for the signal shower \cite{gkkk,Akopdzhanov:1976pr}. It is also used to distinguish hadrons from electrons providing an additional hadron rejection factor of $\simeq 10$ \cite{Davydov:1976vm}.

\item There should be no activity  in the veto counter $V_2$. 
\end{itemize}

\begin{figure*}[tbh!!]
\includegraphics[width=0.33\textwidth]{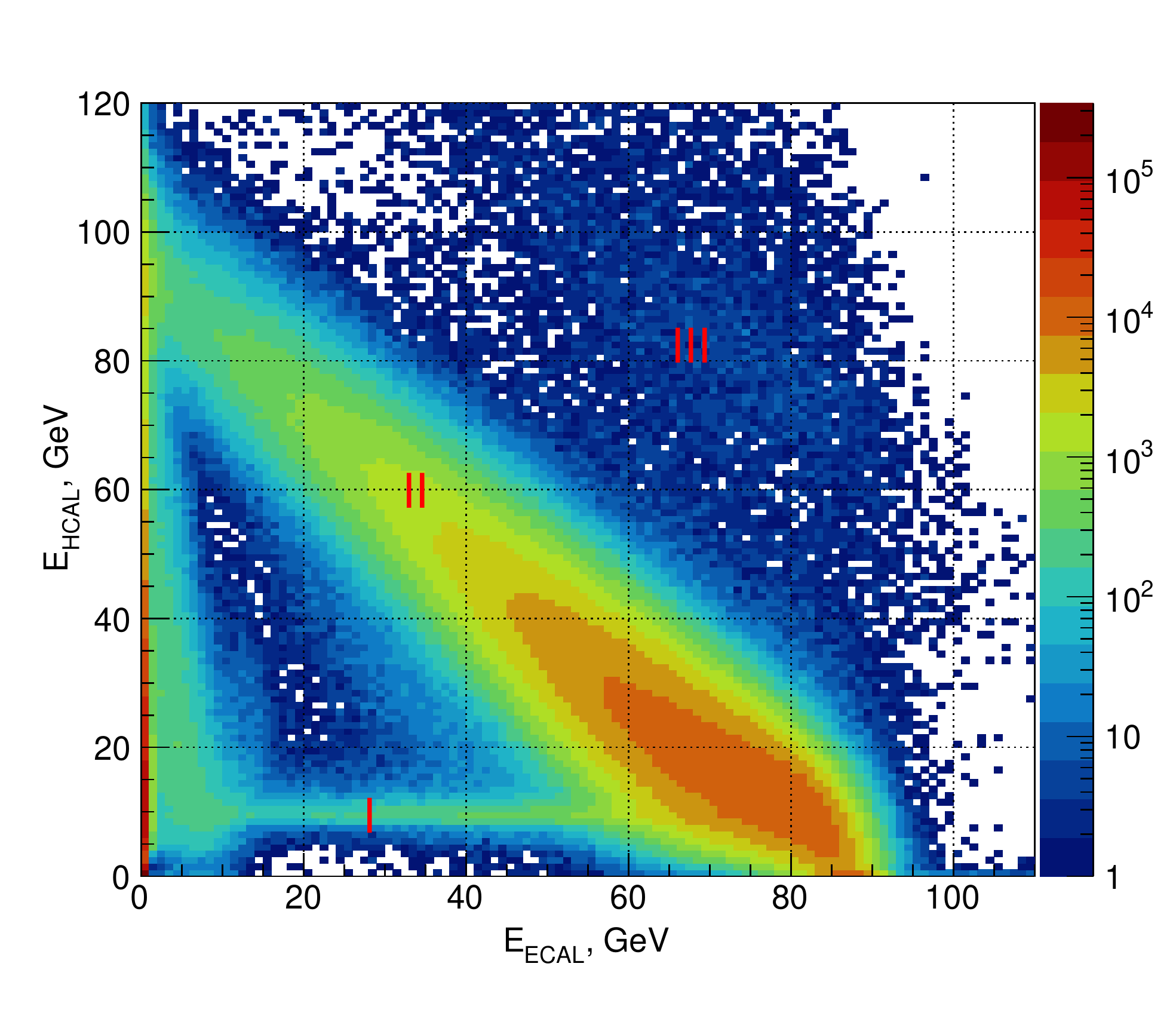}
\includegraphics[width=0.33\textwidth]{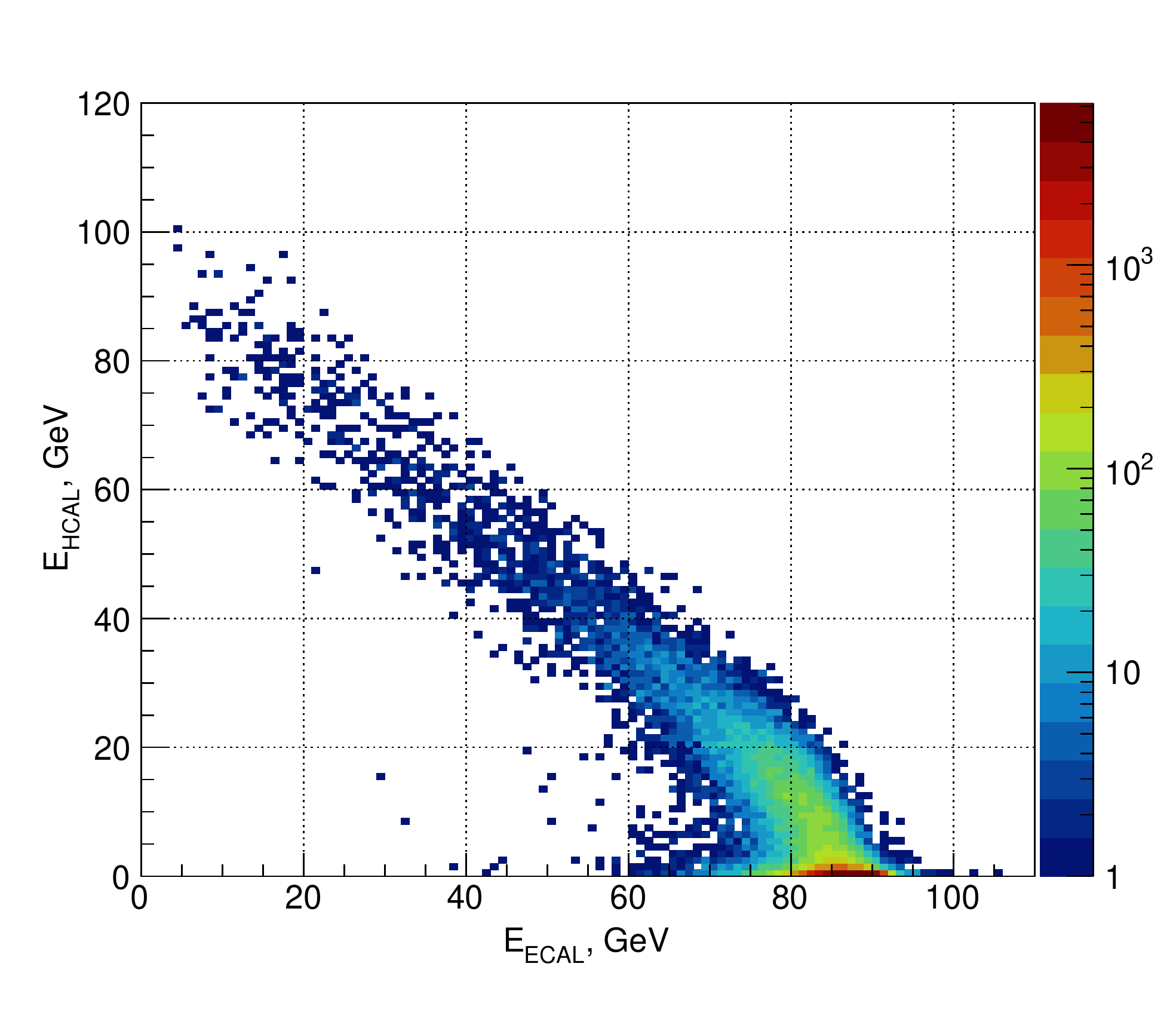}
\includegraphics[width=0.33\textwidth]{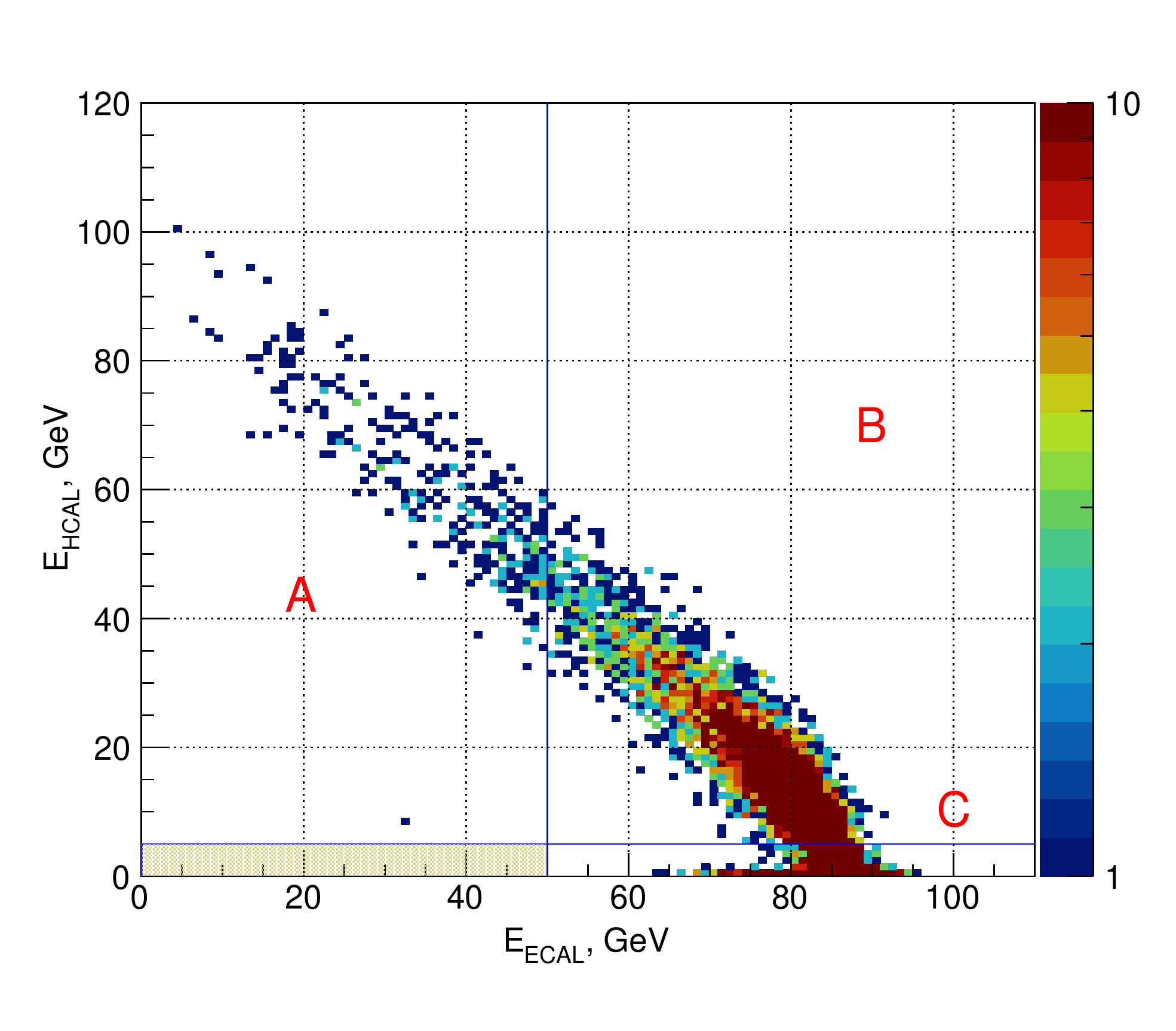}\\
\includegraphics[width=0.33\textwidth]{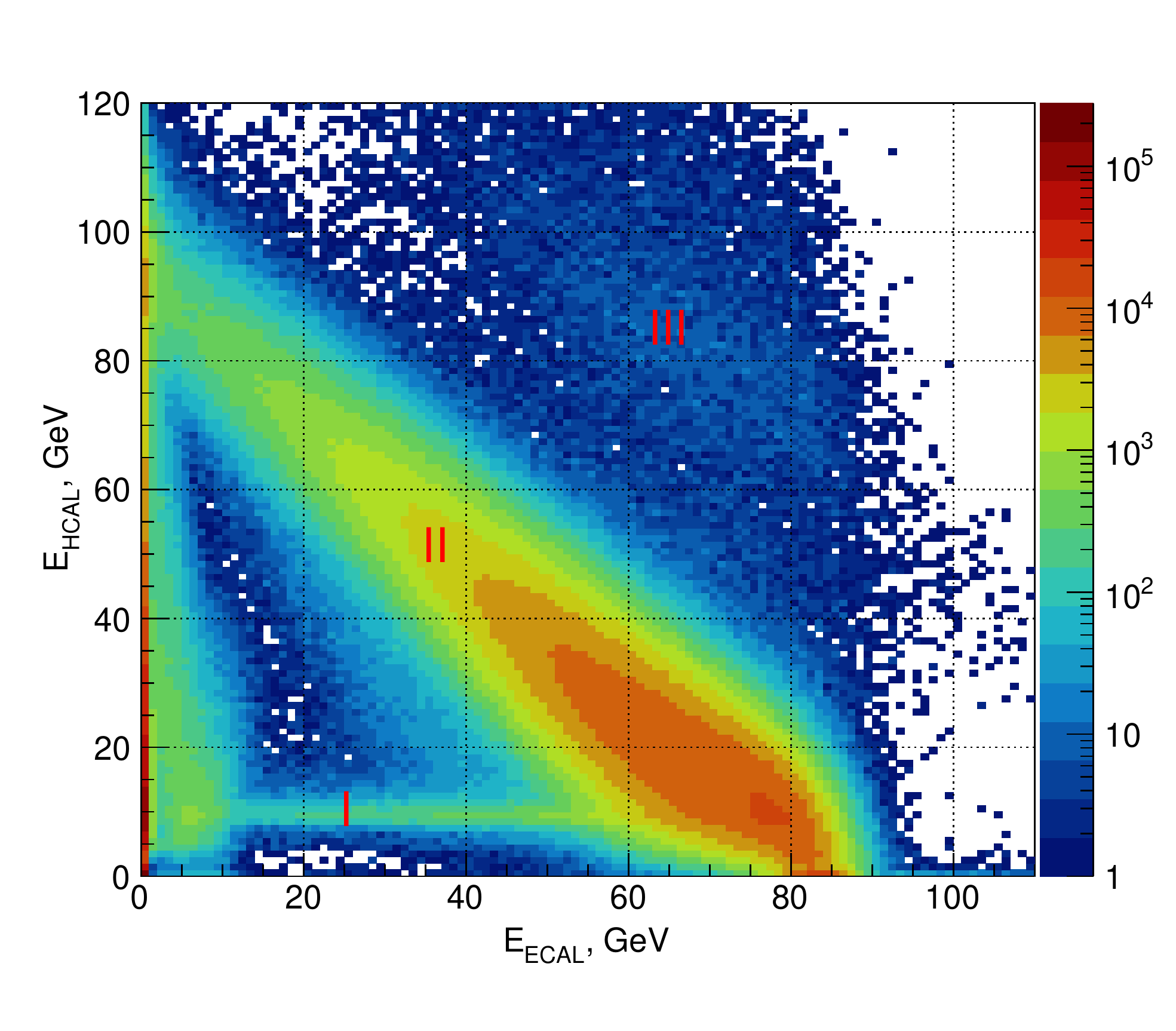}
\includegraphics[width=0.33\textwidth]{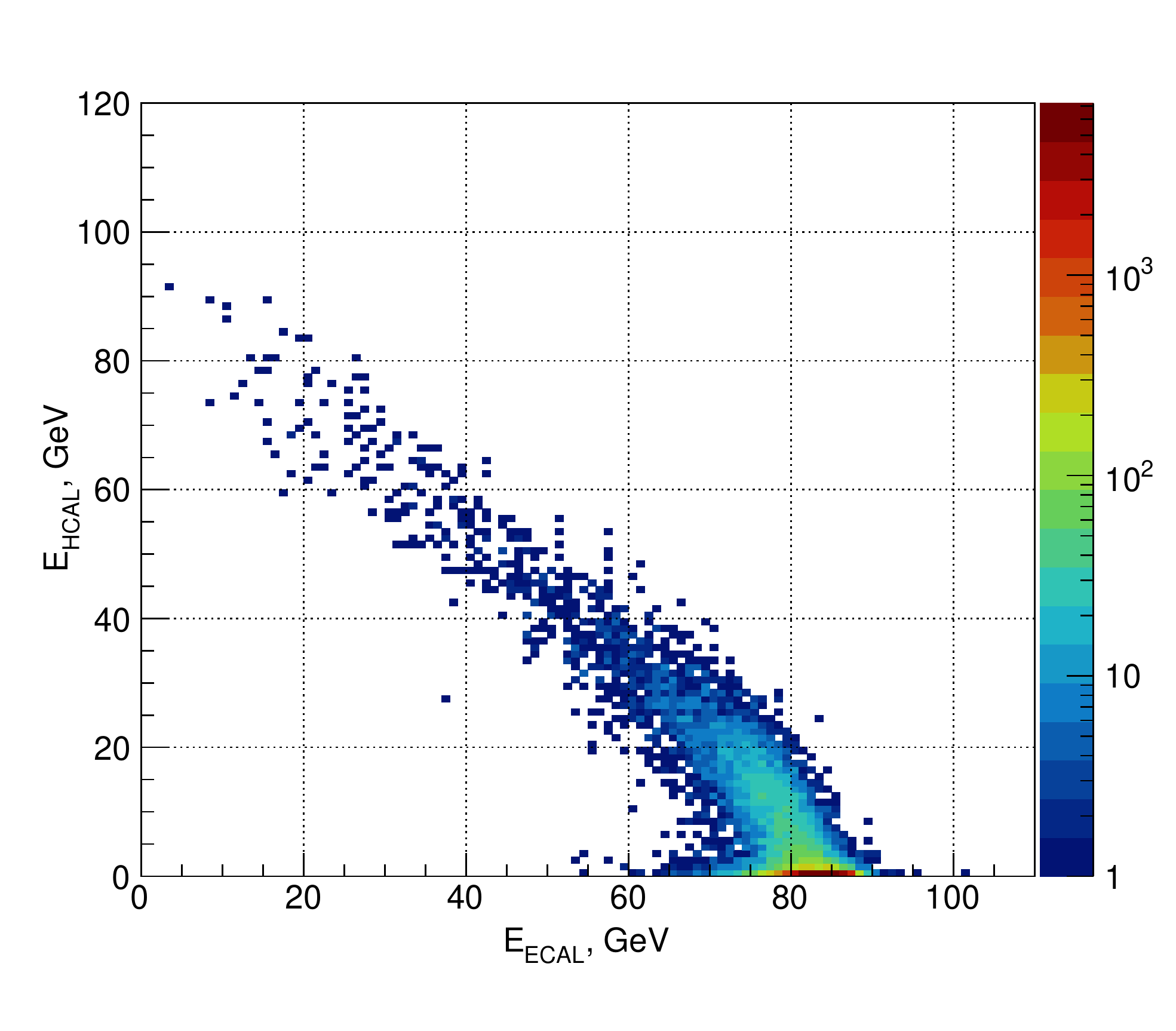}
\includegraphics[width=0.33\textwidth]{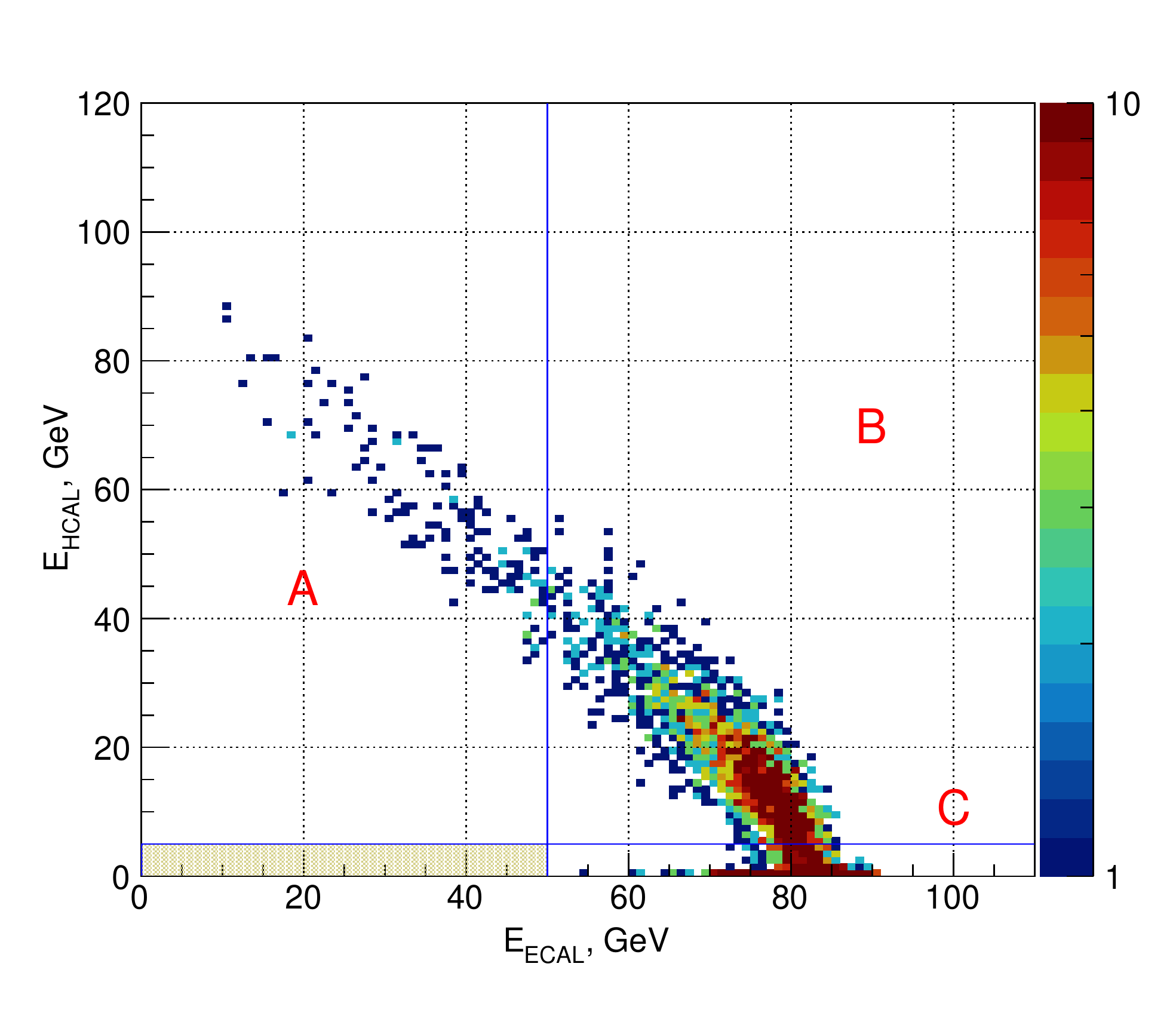} 
\caption{Event distribution in the ($E_{ECAL}$;$E_{HCAL}$) plane from
the runs II(top row) and III (bottom row) data. 
The left panels show the measured  distribution of events at the earlier 
phase of the analysis. Plots in the middle  show the same distribution  after applying  all selection criteria, but the cut against upstream interactions.
The right plots present the final event distributions after all cuts applied.
 The dashed area is the signal box  region  which  is open. 
The side bands A and C are the one  used for the background estimate inside the signal box. For illustration purposes 
the size of the signal box along $E_{HCAL}$-axis is increased by a factor five.
 }
\label{ecvshc}
\end{figure*}

 In total $\simeq 7 \times 10^4 $ events passed these criteria from the combined 2016 data sample.
The final selection is a cut-based and uses the cuts on the ECAL missing energy $E_{miss} = E_{beam} - E_{ECAL}$
and on the energy deposition in the HCAL.
In order to avoid biases in the choice of selection criteria for $A'$ events, a blind analysis was performed, with a preliminary
definition of the signal box as $E_{miss} > 50$ GeV and $E_{HCAL} < 1$ GeV. The HCAL
zero-energy threshold $E^{th}_{HCAL}=1$ GeV in \eqref{sign}, see Sec.\ref{sec:eff} was determined mostly by the noise of the read-out electronics.
Events from the preliminary signal box were excluded from the analysis of the data until the validity of the background estimate in this
region was established.  The cut on $E_{miss}$ was optimized as described in Sec.\ref{sec:results}.
\par In Fig.~\ref{ecvshc} the left panels  show an example of 
the distributions of events from the reaction 
$e^- Z \to anything$ in the  $(E_{ECAL}; E_{HCAL})$ plane measured 
in the runs II(top) and III(bottom) with moderate  selection criteria requiring only  the presence of the SRD tag identifying the beam electrons. 
Here, $E_{HCAL}$ is the sum of the energy deposited in the HCAL1 and HCAL2.
The distributions of events from the run I
with low intensity are similar to the one shown in Ref.\cite{na64prl}.
 Events from the areas I in Fig.~\ref{ecvshc} originate from the QED dimuon production,  dominated by the 
the muon pair photoproduction by  a hard bremsstrahlung photon conversion on a target nucleus:
\begin{equation}
e^- Z \to e^- Z \gamma; \gamma \to \mu^+ \mu^-.
\label{dimu}
\end{equation}
 with some contribution from $\g \g \to \mu^+ \mu^-$ fusion process. 
The $\mu^+\mu^-$ pairs  were characterised by  the HCAL  energy deposition of $ \simeq 10$ GeV. This  rare process whose 
fraction of events with $E_{ECAL}\lesssim 60$ GeV was $\lesssim 10^{-5}$/EOT served  as a benchmark  allowing to verify  the detector performance and as a reference for the background prediction. 
The regions II shows  the events from the SM hadron electroproduction in the ECAL which satisfy 
the energy conservation $E_{ECAL} + E_{HCAL} \simeq 100$ GeV  within the detector energy resolution. 
The leak of these events to the signal region mainly due to  the HCAL  energy resolution   was found to be negligible. 
The fraction of events from the region III was
due to pileup of $e^-$ and  beam hadrons. It was
 beam rate dependent with a typical value  from about   a few $\%$ up to $\simeq$ 20\%.

\section{Dimuon events from the reaction $e^- Z\rightarrow e^- Z \mu^+ \mu^-$}
\label{sec:dimuon}

To evaluate the performance of the setup, a  cross-check between a clean sample of $\gtrsim 10^4$ observed and MC simulated
$\mu^+ \mu^-$ events was made. The process \eqref{dimu} was used as a benchmark allowing to verify the reliability of the MC simulation
and to estimate the corrections to the signal reconstruction efficiency and possible additional uncertainties in the $A'$ yield calculations.  
Let us first briefly review the description of the gamma  conversion into a muon-antimuon pair implemented  in Geant4.
The dimuon production was also used as a reference for the prediction of background, see Sec. \ref{sec:bckg}. 
   
\subsection{Simulation of dimuon events }
The dimuon production has been simulated with  Geant4 \cite{geant} and  a code developed  by NA64 used also for simulation of dark photon production \cite{gkkk}.  Here, we report our comparison with data based  mostly  on Geant4 simulation for decays and propagation of muons through the detectors. However, we anticipate that comparison of dimuon results with the NA64 code will follow in the future, as it will be  an important additional cross-check of the  $A'$ yield  calculations  reported in this work  and in Ref.\cite{kk}.

The gamma conversion into a muon-antimuon pair 
\begin{equation}
\gamma Z \rightarrow \mu^+\mu^- Z
\end{equation}
on nuclei is a well known reaction in particle physics (Bethe-Heitler process).
The simulation of this reaction in Geant4 is based on the differential cross section for electromagnetic creation of muon pairs  
on nuclei $(A, Z)$ in terms of the energy fraction of muons 
\cite{hbur,skel}:
\begin{equation}
\frac{d\sigma}{dx_+} = 4 \alpha Z^2 r^2_{\mu}\Bigl(1 - \frac{4}{3}x_+x_{-}\Bigr) \log(W) \,,
\label{diffdimu}
\end{equation}
where $x_+ = \frac{E_{\mu^+}}{E_{\gamma}}$,  $x_{-} = \frac{E_{\mu^{-}}}{E_{\gamma}}$, 
$\alpha = \frac{1}{137}$ and  
$r_{\mu} = \frac{\alpha}{m_{\mu}}$ is the classical radius of muon and 
\begin{equation}
W = W_{\infty} \frac{1 + ( D_n\sqrt{e} -2) \delta/m_{\mu}}
{1 + BZ^{-1/3}\sqrt{e}\delta/m_e} \,,
\end{equation}
where $ W_{\infty} = \frac{BZ^{-1/3}}{D_n} \frac{m_{\mu}}{m_e} $, $\delta = 
  \frac{m^2_{\mu}}{2E_{\gamma}x_+x_{-}}   $, $\sqrt{e} =1.6487$.
For hydrogen the values $B = 202.4$ and $D_n = 1.49$ are used. 
For other nuclei    those are  $B = 183$ and 
$D_n = 1.54 A^{0.27}$. Here,  $A$ is the atomic number of the nuclei.
The differential cross section  is symmetric in $x_+$ and $x_{-}$, 
and a  relation is 
\begin{equation}
x_+x_{-} = x_{\pm} - x^2_{\pm} \,
\end{equation}
takes place.
The differential cross section  \eqref{diffdimu} can be rewritten in the form
\begin{equation}
\frac{1}{\sigma_0} \frac{d\sigma}{dx} = [1 -\frac{4}{3}(x - x^2)]
\frac{\log W}{\log W_{\infty}} \,,
\end{equation}
Here $ x = x_{+}$ or $x = x_{-}$.
The total cross section was obtained by integration of the differential 
cross section, namely
\begin{equation}
\sigma_{tot}(E_{\gamma}) = \int^{x_{max}}_{x_{min}} \frac{d\sigma}{dx_+}dx_+ \,,
\end{equation}
where $x_{max} = \frac{1}{2} + \sqrt{\frac{1}{4} - \frac{m_{\mu}}{E_{\gamma}}}$, 
$x_{min} = \frac{1}{2} - \sqrt{\frac{1}{4} - \frac{m_{\mu}}{E_{\gamma}}}$.
Numerically for Pb nuclei $ \sigma_{tot} = 30.2;~334;~886 ~\mu$b for $ E_{\gamma} =
1, ~10,~100$ GeV,  respectively.

Note, that formula \eqref{diffdimu} for the cross section was obtained from the 
tree level formula for the differential cross section $\gamma Z 
\rightarrow  \mu^+\mu^-Z$  by taking into account  both the  atomic and nuclear form-factors and 
 without using  the WW approximation of equivalent photons. Even though the 
 production mechanisms of the $A'$ and $\mu^+ \mu^-$ pair are  different, the number  of $A'$ and dimuon events,  
are both proportional to the square of the Pb  nuclear form factor $F(q^2)$ and are sensitive to its shape. 
As the  mass $(\ma \simeq m_\mu)$ and $q^2$ $(q\simeq \ma^2 /E_{A'}\simeq m_\mu^2 /E_{\mu})$ ranges of the final state   for both reactions are similar, the observed difference can be considered  as due to the accuracy of the  dimuon
yield  calculation for  heavy nuclei.
\begin{figure}[tbh!]
\includegraphics[width=0.5\textwidth]{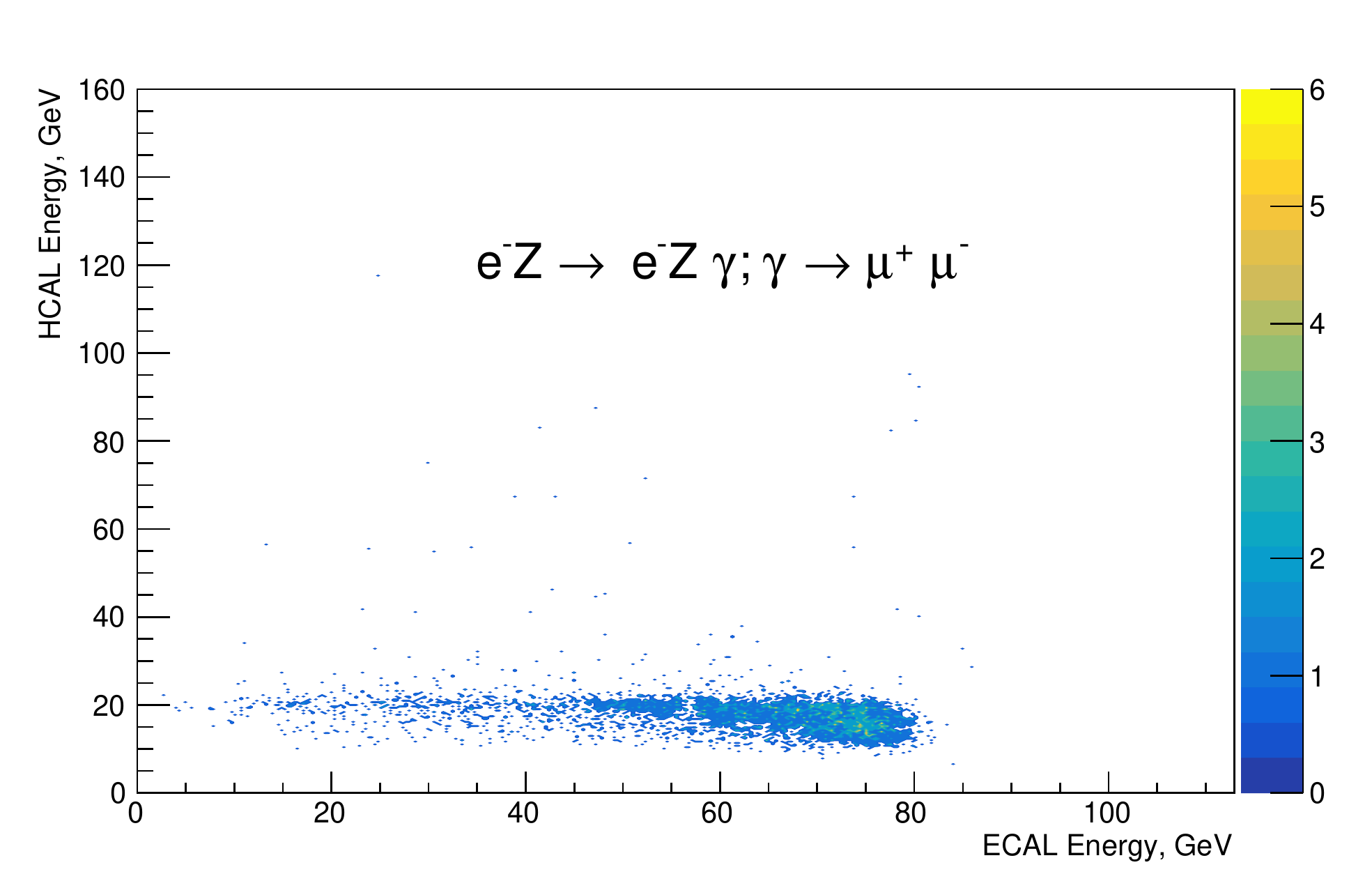}
\caption{Selected  dimuon events in the  $(E_{ECAL};E_{HCAL})$ plane.}
\label{dimuon}
\end{figure}   

\subsection{Yield of dimuon  events}
The dimuon events were selected with the trigger \eqref{trigger}, which 
accepted only events with the ECAL energy deposition
smaller than $\simeq$ 80 GeV. Because only  muons can 
punchthrough the total length of  the  modules HCAL1-3 ($\simeq 21\lambda_{int}$)   without interactions,  the  selection  was  based on the   requirement of the energy deposited  in   HCAL1   and HCAL4  modules to be  in the range 
 $1 \lesssim E_{HCAL1,4} \lesssim 6 $ GeV,  which  is comparable with that  expected from a single muon or dimuon
pair.  In Fig.~\ref{dimuon} the distribution of selected dimuon events in the 
($E_{ECAL};E_{HCAL}$) plane is shown. Here, the HCAL energy is defined as the total energy deposited in the four HCAL modules. 
\begin{figure*}[tbh!!]
\includegraphics[width=0.5\textwidth]{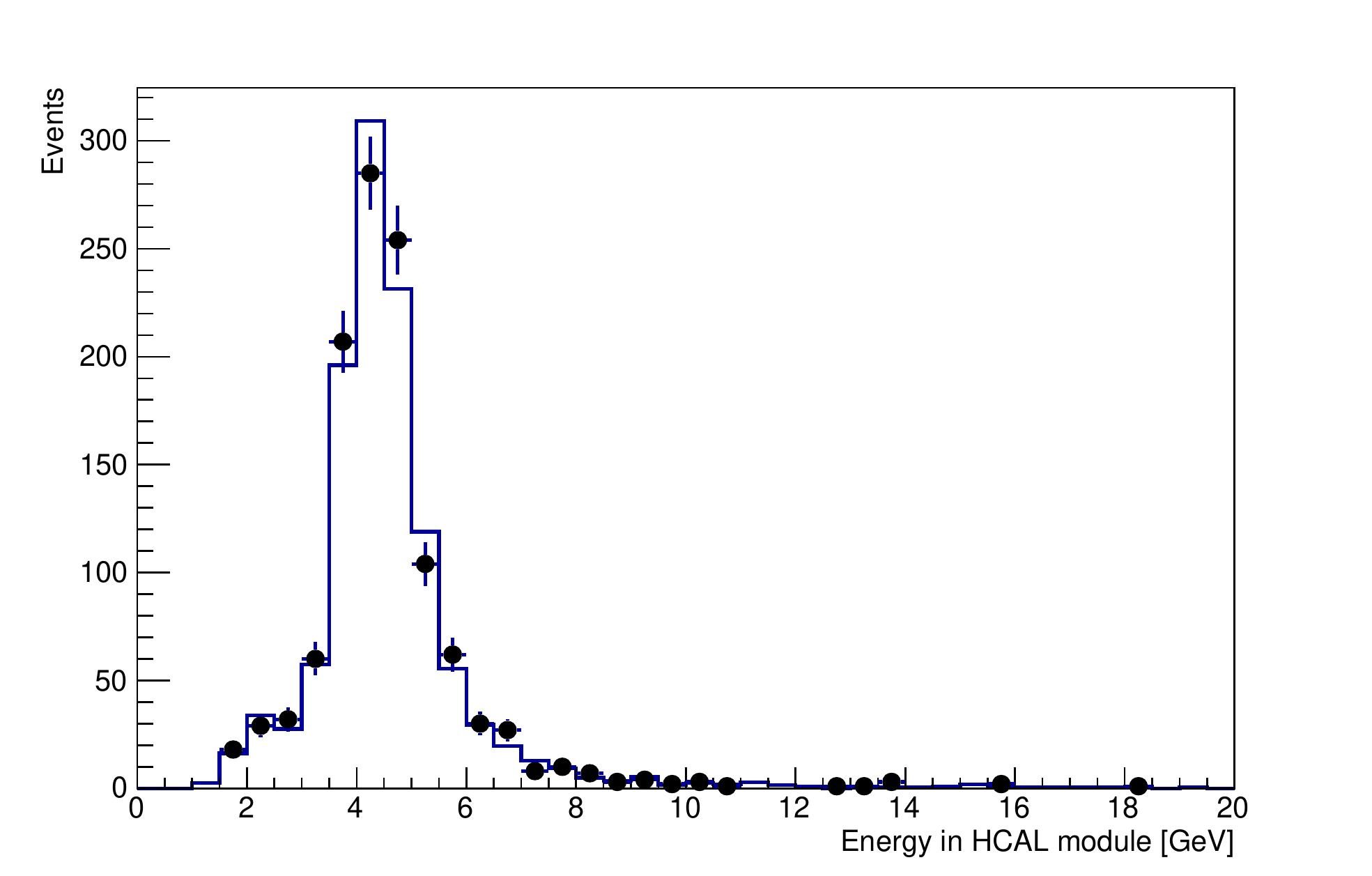}
\hspace{-0.cm}{\includegraphics[width=0.5\textwidth]{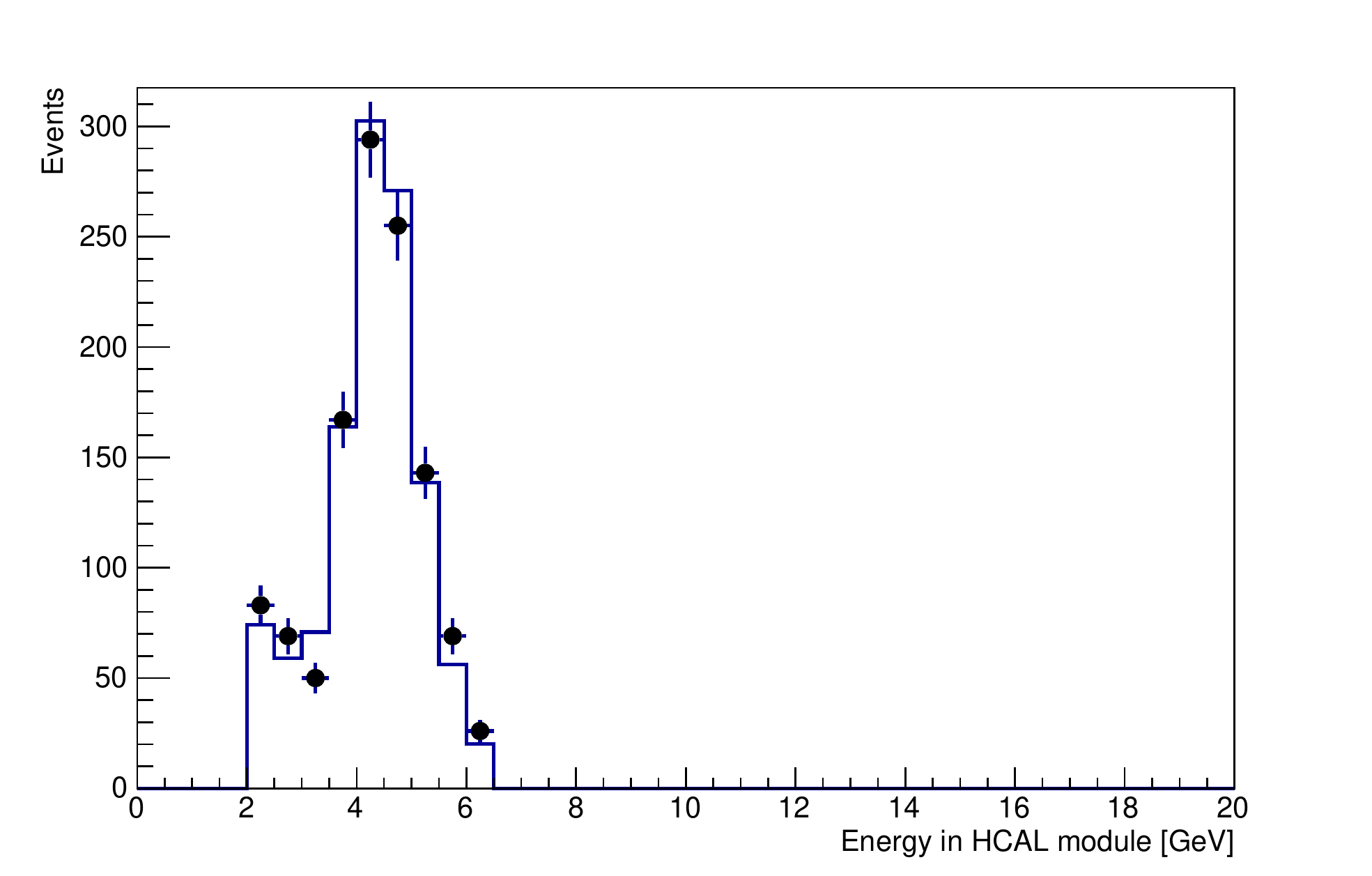}}
\caption{ Comparison of  expected (solid) and measured (dots) distributions of dimuon events in the   HCAL2 (left panel)  and 
HCAL module 3 (right panel).  The  small bump at $\simeq2.5$ GeV  originates  from a single muon of the pair when the other one stops 
in the previous module. The spectra are normalised to the same number of events. }
\label{dimuhcal23}
\end{figure*}   

The dimuon yield was estimated  from the observed number of  reconstructed  dimuon events.  
The comparison of the number of observed ($n_{2\mu}^{data}$ ) and predicted ($n_{2\mu}^{MC}$ ) $\mm$ pairs and 
the corresponding reconstruction efficiency ($\frac{n_{2\mu}^{data}}{n_{2\mu}^{MC}}$) is shown in Table \ref{tab:dimuoneff}.  One can see, that
the reconstruction efficiency of $\mm$ pairs  were found to be  beam rate dependent. 
\begin{table}[hbt]
\caption{Dimuon selection efficiency for the data samples from the runs I-III  obtained at  different beam intensity  for $E_{ECAL} <  60$ GeV}   \label{tab:dimuoneff}
\begin{center}
\begin{tabular}{|c|c|c|c|c|c|}
\hline
Data  & beam  && &  & Efficiency \\
 sample & intensity, $10^6$ &~ $n_{EOT}$, $10^6$ ~& ~$n_{2\mu}^{MC}$ ~ & ~$n_{2\mu}^{data}$~ & reduction \\
 &&&&&factor $f$\\
\hline
\hline
run I & 1.8& 171& 1223& 1124& 0.92  \\
\hline
run II&3.2& 208.5& 1491& 1268& 0.85\\
\hline
run III& 4.6& 597& 4271& 3417& 0.81\\
\hline
\hline
\end{tabular}
\end{center}
\end{table}
The difference between  the number of  observed  and MC predicted $\mu^+ \mu^-$ events  with $E_{ECAL} \lesssim 60$ GeV  is  the range 8-19\%. It   can be interpreted as due to the inaccuracy of the  dimuon yield  determination  for  heavy nuclei target and, can be conservatively  accounted for as  an additional 
 reduction factor $f$  of the signal  efficiency,  which depends on the beam intensity. The uncertainty in this factor includes uncertainty due to the 
difference of  the ECAL energy spectra for dimuon and $A'$ events which is taken into account by the reweighting procedure  discussed below in Sec.\ref{sec:eff}.  
 
\subsection{The HCAL and ECAL energy distributions}

\begin{figure}[tbh!]
\includegraphics[width=0.5\textwidth]{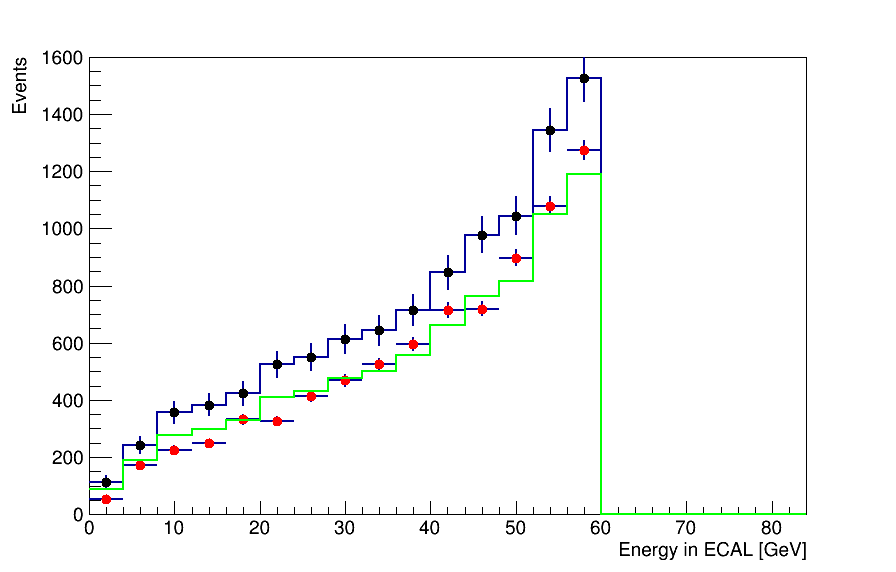}
\caption{Distribution of energy deposited in the ECAL target by the 
scattered electron from the reaction \eqref{dimu} 
 for the selected dimuon events from the data sample of  the   run III (red points) and MC events (green histogram). Spectra are normalized to the same number of events. The unnormalized MC distribution (top histogram) is also shown  with corresponding errors for the each bin.}
\label{invis_dimu2351-9}
\end{figure}

An example illustrating  good agreement between  distributions  of energy deposited by $\mm$ in the HCAL module 2,  for the data and MC 
  is shown in Fig.~\ref{dimuhcal23}. On  the right panel of the plot one can see
a small peak at $\simeq 2.5$ GeV from  single  muons originated from events when one of the muon from the $\mm$ pair did not reach the HCAL3. 
An additional cross-check was made by comparing the distributions of the energy
 $E_{ECAL}$  deposited by scattered electrons from the reaction \eqref{dimu} in the ECAL taking into account small corrections due to dimuon energy depositions.  
 This comparison of the data vs MC  $E_{ECAL}$ distributions for the high intensity run III  is shown in 
Fig.~\ref{invis_dimu2351-9}.   One can see
 that the predicted and measured spectra are in a reasonable agreement  and  are not significantly distorted by  pileup events. 

\section{Signal efficiency}
\label{sec:eff}
 Several signal detection efficiency  contribute  to the value of  
$\epsilon_{tot}(\ma)$  in the NA64  detector:
\begin{equation}
\epsilon_{tot}(\ma)  = \epsilon_{e}\cdot \epsilon_{A'}\cdot \epsilon_{ECAL}\cdot \epsilon_{V}\cdot \epsilon_{HCAL} 
\label{eff}
\end{equation}
where  $\epsilon_{e},~ \epsilon_{A'},~  \epsilon_{ECAL},~  \epsilon_{V}$ and $ \epsilon_{HCAL}$ are the efficiency  factors  for the primary $e^-$ selection,
which include also the reduction factor $f$ discussed in Sec. V.B,   the $A'$ acceptance in  the signal box range, and the efficiencies 
 for the signal to
pass the ECAL, $V_2$ , and HCAL selection criteria, respectively. The  $\epsilon_{ECAL}$ value includes also the ECAL spectrum reweighting factor
discussed below in Sec. VI.A.  
 These  factors  were determined from the 
sample obtained with MC simulations and from the data samples of $e^-$ and 
dimuon events. 
The flux and spectra of the $A'$s produced in the ECAL target
by primary electrons  were calculated using the 
approach reported in ref. \cite{gkkk}  taking into account  the development 
of the signal e-m shower from reaction 
\eqref{e-a}  in the ECAL target (see, Sec. V).  

\subsection{The ECAL signal efficiency}
The reconstruction efficiency $\epsilon_{ECAL}$ for  signal events  was  calculated  for different $A'$ masses as a function of  energy deposited in the ECAL.
  Compared to the ordinary e-m shower,  the $\epsilon_{ECAL}$  value  for the e-m  shower induced by an $A'$ event 
 has to be corrected  mainly due to difference in the longitudinal  e-m showers development   at  the early stages in the PS detector \cite{gkkk}. 
This correction   depends  on the  threshold  $E^{th}_{PS}$  of the energy deposited in the PS used in the trigger  \eqref{trigger} and  was  typically 
$\lesssim (5\pm 3)\%$ where the errors came from the $E^{th}_{PS}$  threshold variation during data taking.

The sensitivity of the NA64 experiment is defined by  the number of 
accumulated events which depends on the beam intensity. The intensity is limited by the pulse  duration ($\tau_{ECAL} \simeq 100$ ns) from the ECAL MSADC shaper resulting in a maximally allowed electron counting rate  of $ \simeq  10^{6}~ e^- /$s in order to avoid significant loss of the signal efficiency 
due to the pileup effect. To evade this limitation,  we have  implemented  a pileup removal algorithm to allow for high-efficiency  reconstruction of the $A'$ signal  and energy in high electron pileup environments,  and run the experiment at the electron beam rate $\simeq$ a few $10^6~e^-$/spill.
 This is in particular important in the case
of signal events, because  the shape of the $E_{ECAL}$ spectrum can be used for the $A'$ mass evaluation \cite{gkkk}. 
The shape is in particular sensitive to the mass in the low  energy 
region  which is the most affected by the pileup pulses which may occur somewhat earlier or later than 
the desired pulse and may seriously affect the reconstruction efficiency of signal  events. 

A simple pileup removal algorithm was used in the analysis of the data and MC samples of  
events  obtained  for  high beam intensity.  All ECAL cells 
were  requested to have a single MSADS peak with a cell-time within $\pm 2$~ns of the trigger time  if the energy deposited in the cell was  more than 1 GeV. 
If  several peaks were found, the one closer to the expected cell-time 
position was selected, and an attempt was made to remove contributions from the 
neighbouring  pileup peak(s) to the signal area. The efficiency of the pileup removal  
algorithm as a function of the $E_{ECAL}$ value  was  estimated  by 
 using  clean data and MC   dimuon samples obtained at different 
intensities as described below. This  method  was also used to evaluate  the efficiency of the $A'$ signal 
reconstruction in the energy range predicted by the simulations. 
\begin{figure}[tbh!!]
\centering
\includegraphics[width=0.5\textwidth]{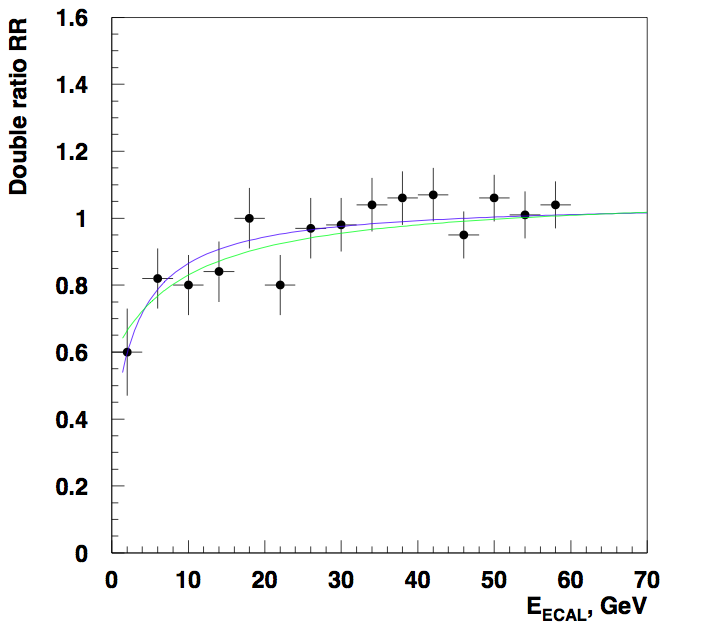}
\caption{Double ratio $RR$ as a function of the ECAL energy. The color curves  represent an example of the empirical fitting  functions. }
\label{fig:d-ratio}
\end{figure}   
At high intensity the dependency of dimuon events reconstruction efficiency on $E_{ECAL}$ can be important.
For this reason using the efficiency corrections directly from the overall ratio data/MC in dimuons can be inaccurate
due to the difference in $E_{ECAL}$ spectra. In order to check this the samples of reconstructed $\mm$ events
from data and MC simulation were compared with more details. For this purpose the individual correction $RR$ factors, ratios between efficiency measured in data and in Monte Carlo, have been derived and they were applied as an event weights in the MC simulations to obtain a better 
 agreement between simulated and real data samples. These scale factors are used to correct Monte Carlo efficiencies to agree with measurements on data samples. 
 The $RR$ ratios have been formed as a function of $E_{ECAL}$ value:
\begin{equation}
R_{Data (MC)} =\big( \frac{n^{EC}_{i}}
{n^{EC}_{tot}}\big)_{Data (MC)}
\label{d-ratio}   
\end{equation}
where $n^{EC}_i$ and $n^{EC}_{tot}$ is the number of events in i-th bin and the total number of events in the ECAL energy distribution, see 
Fig.~\ref{invis_dimu2351-9}. The ratio of the above ratios  $RR=R_{Data}/R_{MC}$ 
is then a measure of any additional differences in the signal reconstruction efficiency
between data and MC as a function of the energy deposited in the ECAL. In Fig.~\ref{fig:d-ratio}
the distribution of  $RR$ values  over the ECAL energy range from 1 to 60 GeV 
is shown for the beam intensity $\simeq 5\times10^6~e^-$/spill from run III. 
\begin{figure*}[tbh!]
\includegraphics[width=.45\textwidth]{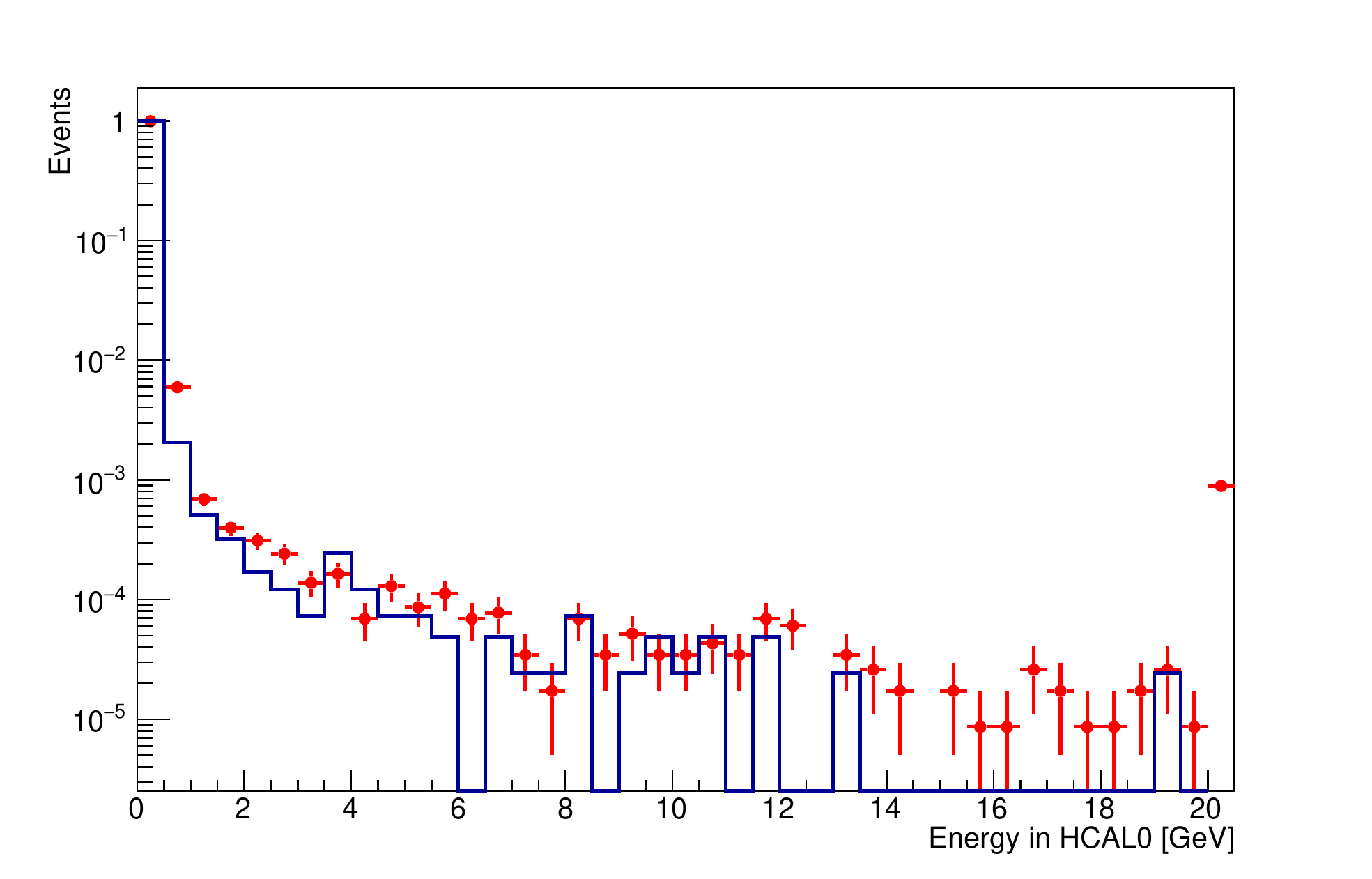}
\includegraphics[width=0.45\textwidth, height=0.27\textwidth]{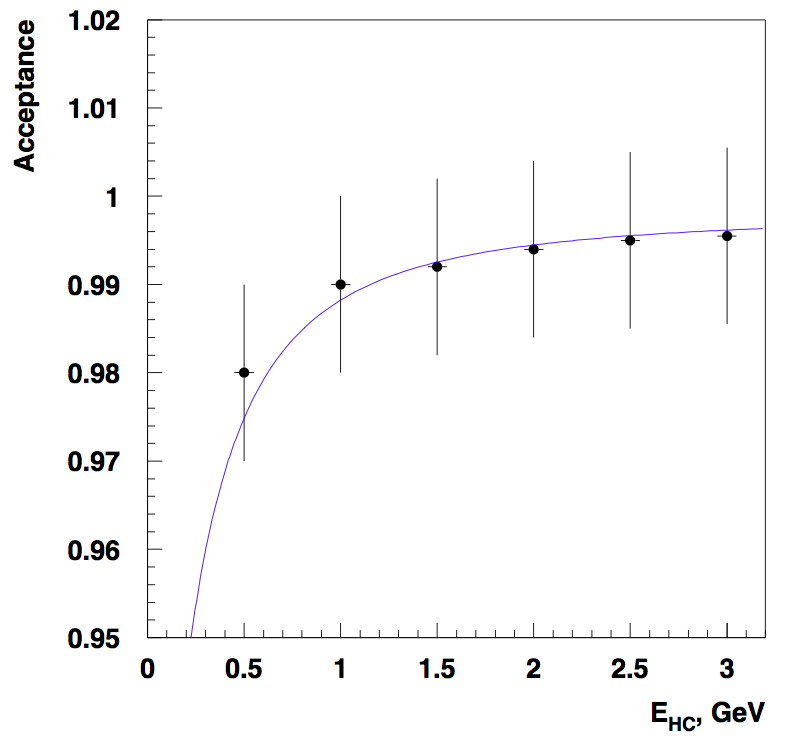}
\caption{The left hand side panel shows distribution of the leak energy from the ECAL to the HCAL from the 100 GeV e-. The right hand side panel represents 
the 100 GeV $e^-$ detection efficiencies  as a function of the HCAL energy threshold.}
\label{hcalleak}
\end{figure*}
One can see that the  method works well for the  energy region, $E_{ECAL}\gtrsim  5~\rm GeV$, 
when the distortion of the ECAL spectrum from the pileup effect  is relatively small at any  distance between the true and pileup pulses in the ECAL 
yielding a correction factor close to 1. In the low energy region $E_{ECAL}\lesssim 10~\rm GeV$, the reconstruction efficiency 
is more affected by the statistical uncertainties in the shape of the reconstructed 
pileup pulse, and the true pulse is not identified well anymore.\
This difference was used to additionally correct the MC efficiency for the signal in order to
account for pileup and other effects not present in the MC. 
For this purpose  the signal spectrum shown in Fig.~\ref{ecal-spectra}
was reweighted  by applying bin-by-bin corrections to the signal  efficiency for the given energy  obtained from 
dimuon data sample as shown in Figs.\ref{invis_dimu2351-9} and \ref{fig:d-ratio}. 
This procedure   results in the  overall correction factor to the signal efficiency $0.93$  for the case of highest intensity of the run III, 
and is $\gtrsim 95\%$ for the runs I and II.  It  is slightly  mass dependent.  

\subsection{The Veto and HCAL cuts and efficiency corrections.}
The $V_2$ and HCAL signal acceptance  was  defined as a fraction of events below the corresponding zero-energy cuts: 
\begin{equation}
\epsilon_{HCAL(V_2)} = \frac{n_{HCAL(V_2)}(E<E^{th}_{HCAL(V_2)})}{n_{tot}}
\end{equation}
where $n_{HCAL(V_2)}(E<E_{th})$, $n_{tot}$ is the number of events below the threshold
energy $E_{th}$ and the total number of events, respectively.
The shape of the distributions
of energy deposited in these detectors from the leak of the signal shower energy, deposited in the ECAL, was simulated for different
$A'$ masses and cross-checked with measurements at the $e^- $ beam of several energies.
The HCAL energy cut was chosen by using the distribution of $E_{HCAL}$ in the data from the calibration runs,
where the ECAL threshold was removed from the  trigger. The admixture of non-electrons in these runs can be neglected.
In Fig.~\ref{hcalleak} the  right  panel shows the dependence of $\epsilon_{HCAL}$ on the
energy threshold $E_{HCAL}^{th}$.
One can see that for the value $E_{HCAL}^{th}\simeq 1 $ GeV the
acceptance is $\epsilon_{HCAL}\simeq 0.98$ for the maximal beam intensity.

\par The corrections to efficiency  $\epsilon_{V}$ and $\epsilon_{HCAL}$ in the $V_2$ and HCAL 
were determined directly from the data using the calibration runs.
The left panel  in  Fig.~\ref{hcalleak} shows  
an example of the measured distribution of the energy in the HCAL in such run.
In general, the agreement with MC spectrum is good. The small differences between data and MC distributions are dominated by the pileup effects.
The results for the $V_2$ acceptance look quite  similar.
Quantitatively, to estimate the $V_2$  and HCAL signal efficiency  we studied the ratios of the number of events above and below the
corresponding cuts  used for the $A'$  event selection.

An example of the summary of corrected efficiencies  used for calculations 
of the limits for the run III data sample obtained  with a beam intensity 
$\simeq 5\times 10^6~e^-$/spill
is presented in Table \ref{tab:accept}. 
These efficiencies were slightly different for the  data samples from runs I and II,  mostly  because 
of the different pileup algorithm efficiencies which were rate  dependent and 
also determined from measurements in calibration runs at different  beam rates. 
The quantities $\epsilon_e, \epsilon_{ECAL}$ were the most rate-dependent 
detection efficiencies of the tracker chambers and SRD, and ECAL cluster reconstruction, respectively.
The DAQ  deadtime was a function of the 
beam rate and was 7.4\% averaged over the full data-taking period.
\begin{table}[tbh!!]
\begin{center}
\begin{tabular}{|c|c|c|}
\hline
\hline
 item &  Efficiency & sample    \\
\hline
primary $e^-$, $\epsilon_{e}$  & $0.58$ & Data, Dimuons\\
ECAL,  $\epsilon_{ECAL}$ & $0.93 (0.90)$ & Data, Dimuons \\
$V_2$, $\epsilon_{V}$  & $0.94$& Data, MC\\
HCAL, $\epsilon_{HCAL}$  & $0.98$ & Data, MC \\
\hline 
Total & $0.50(0.48)$  &  \\
\hline
\hline
\end{tabular}
\end{center}
\caption{Summary of efficiencies for the signal event selection for the mass $m_{A'}= 10 (100)$ MeV 
 in the data sample obtained for the high intensity run III. For discussion of corresponding uncertainties, see Sec.\ref{sec:syst}.}
 \label{tab:accept}
\end{table}

\par The total number of collected EOT in 2016 was obtained from the recorded  number of events  from the
e-m $e^- Z$ interactions in the ECAL target by taking into account the trigger suppression factor
($\gtrsim 10^2$) and DAQ dead time which was beam rate  dependent. The $e^-$ beam loss due to interactions with the beam line  materials was estimated
to be  small. The trigger and SRD efficiency obtained by using unbiased  samples of events that bypass the selection criteria was found
to be $\simeq 0.95$  and 
$\simeq 0.97$ with a small uncertainty 2\%.
The probability of $A'$ events to pass all selection criteria, $\epsilon_{A'}$  was evaluated by processing the simulated signal events through the same reconstruction program as data, with the same cuts. The $A'$ yield calculated in accordance with  Ref.\cite{gkkk, kk} was corrected for the production cross section as described in Sec.\ref{sec:method}. The overall signal  efficiency   was  in the range  $\epsilon_{tot}\simeq (0.7-0.5)$  decreasing for the  higher intensity run.

\section{Systematic uncertainties}
\label{sec:syst}
The systematic uncertainties are determined to stem from the overall normalization, signal cross section computations, reweighting the EAL signal energy distribution, description of the dimuon spectrum, and the uncertainty in the signal efficiencies in the Veto and HCAL. Systematic uncertainties are determined by varying cuts and taking the largest change in the calculated rate as the systematic error.  Details of the systematic checks are given below.
\par The  $10\%$  additional uncertainty, estimated from the 
comparison of the cross sections calculated  in  \cite{Liu:2017htz} and \cite{kk} as discussed in Sec. \ref{sec:method}, 
was  taken into account as the systematic error  for the $A'$ production   in the  target. 
Note, that possible  contributions  from  the purity of the target ($\gtrsim 99.9\%$) and $\simeq 22\%$ admixture of spin1/2 isotope $^{207}$Pb  are estimated  to be small.     
Another contributions are  due to the events selection in the ECAL and the  reweighting  procedure of the  ECAL signal spectrum   described in Sec. \ref{sec:dimuon}.
The former was estimated  as a difference in the signal yield with respect to the nominal value due to the PS energy threshold variation during the run. 
This contribution increases for large $A'$ masses.  The 
systematic uncertainty from the reweighting  procedure was  estimated   by varying the parameters of the empirical fitting functions shown in Fig.~\ref{fig:d-ratio}
 and considering differences in the number of obtained signal events. In this  case the quoted systematic uncertainty is taken to be the quadratic sum of the observed shift and the statistical error  of 4\%  on the shift. The reweighting correction is  slightly   $A'$ mass  dependent and has  the  total uncertainty  7\% for the highest intensity run III.  
Other contributions to the systematic uncertainty on the $A'$ signal efficiency come from the choice of the cut threshold and 
definition of the signal  efficiency for the $V_2$ and HCAL, see Sec.\ref{sec:eff}. 
The uncertainties of the corrections for the $V_2$ and HCAL signal efficiency were studied by varying the corresponding 
energy threshold within  the range determined from the data. The calculated variations were   assigning to the systematic errors, which were 
 estimated to be  3\% and 2\% for the $V_2$ and HCAL, respectively. 
 An example of sources and the corresponding  magnitudes of the systematic errors for the $A'$ masses 10 and 100 MeV, estimated  for the run III is shown 
  in Table \ref{tab:syst}.  
\begin{table}[tbh!] 
\begin{center}
\caption{Summary of  systematic uncertainties  for  the mass $m_{A'}= 10 (100)$ MeV in  the high intensity run III.}\label{tab:syst}
\vspace{0.15cm}
\begin{tabular}{lr}
\hline
\hline
Source of the error& Estimated  error\\
\hline
&\\
Normalization & \\
\hline 
number of collected EOT, $n_{EOT}$  &  2 \% \\
\hline 
&\\
$A'$ Yield  & \\
\hline 
signal cross section & $10\%$ \\
\hline 
&\\
$A'$ efficiency & \\ 
\hline 
primary $e^-$ selection &  4 \% \\
ECAL selection  & 2\% (3.5\%)\\
ECAL  spectrum reweighting & 7 \% (5\%) \\
$V_2$ cut threshold & 3 \% \\
HCAL  cut threshold    & 2 \% \\
\hline 
Total &    9 \%(8\%)  \\
\hline 
\hline
\end{tabular}
\end{center}
\end{table}

\section{Background}
\label{sec:bckg}
\par   The  search for the $\ainv$ decays  requires particular attention to backgrounds, because 
every process  with a single track and an  e-m cluster in the ECAL  can  potentially fake the signal. In this Section we consider all background  sources,  which were also partly studied  in Refs. \cite{Gninenko:2013rka,Andreas:2013lya}.

There are several backgrounds resulting in  the signature of Eq.(\ref{sign}) which can be classified as being due to detector-, physical- and  beam-related sources. 
The selection cuts  to reject these backgrounds have been chosen such that they do not affect the shape of the true $E_{miss}$ spectrum.
\par The estimation of background levels and the calculation of signal
acceptance were both based on the MC simulation, as well as direct measurement 
with the beam.  Because of the small $A'$  coupling strength value,  performing a  complete  detector simulation in order  to investigate these backgrounds down to the level  of a single event sensitivity $ \lesssim 10^{-11}$  would require  a very large amount of computing  time.  Consequently, we have estimated  with MC simulations all known backgrounds to the extent that it is possible. Events from particle interactions or decays in the beam line, pileup activity created from them,  hadron punchthrough  from the target and the HCAL were  included in the simulation of background events.  Small event-number  backgrounds such as the decays of the beam  $\mu, \pi, K$ or $\mu$s from the reaction of dimuon production were  simulated with the full statistics  of the data.     Large event-number processes, e.g.   from $e^-$ interactions in the target or beam line,  punchthrough  of secondary hadrons were  also studied, although simulated samples with statistics  comparable to the data were not feasible. 
To eliminate  possible instrumental effects not present in the MC calculations, the uniformity scan of the  central part of the ECAL target was performed with $e^-$ by using the MM3 and MM4. We also examined the number of events observed in several regions around the signal box, which  were statistically consistent with the estimates. 

The main detector background sources are related to 
\begin{itemize}
\item {\it Instrumental effects}. 
The leak of energy throughout the possible holes, cracks, etc.
in the  downstream coverage  of the detector which allows secondary particles to pass through without 
interactions.  To study this effect a $X-Y$ scan over the transverse area of the ECAL and HCAL detector 
has been performed with a particular attention to the boundaries between cells, fibers positions, and  dead materials. 
No significant leak of energy has been observed.    

\item {\it Detector hermeticity}.
The fake signature~ of Eq.\eqref{sign} could also arise when either: i) a high-energy  bremsstrahlung 
photon from the reaction $eZ\to eZ\g$, or ii) leading hadron $h$ from the reaction $ e Z\to e Z X + h$    in the target escape detection  due to punchthrough in 
the HCAL.  The reaction i) may occur if an energetic photon  induces a photo-nuclear reaction accompanied by the emission of a leading neutral particle(s), such as e.g. a neutron. The neutron then could be undetected in the rest of the detector. Taking into account the estimated non-hermeticity of the detector, the probability of the  reaction  is found to be  $\lesssim 10^{-14}$.   
   For the charged secondaries the punchthrough is highly suppressed by the 
observation of energy deposition in the HCAL modules. As the number of photoelectrons per MIP crossing the HCAL module was measured to be  in the range  
$n_{ph.e.}  \simeq 150-200 /$ MIP the inefficiency of 
the punchtrough detection is $\lesssim 10^{-10}$  making the overall background negligible.  
\par 
For the case ii) the  punchthrough  probability of a leading neutral hadron, such as a neutron and/or $K^0_L$, is defined by $\exp(-L_{tot}/\lambda_{int})$,
where $L_{tot}$ is the (ECAL+HCAL) length sum. It has been estimated  separately  with a pion beam and compared with simulations
\cite{Andreas:2013lya} .
  It has been found that the overall hadron
punchthrough probability  is below  $10^{-12}$ for the total thickness of the ECAL and HCAL of about 30 $\lambda_{int}$. This value  should be multiplied by a  factor $\lesssim 10^{-4}$, which is the probability of a   leading hadron electroproduction in the ECAL target. Taking this into account  the final estimate results to
  the negligible level of this background per incoming electron. 
The HCAL non-hermeticity for high energy neutral hadrons was cross-checked  
with Geant4-based MC simulations \cite{gkkk}. 
For the energy threshold $E^{th}_{HCAL} \simeq 1$ GeV the non-hermeticity is expected to be at the level $\lesssim 10^{-9}$.  Taking into account the probability to produce a single leading hadron per incoming electron as  $P_{h} \lesssim 10^{-4}$,  an overall level of this background of $\lesssim10^{-13}$ is obtained. 
This  is in agreement with the above  rough estimate. 
\begin{figure}[tbh!!]
\centering
\vspace{-.0cm}{\includegraphics[width=0.5\textwidth]{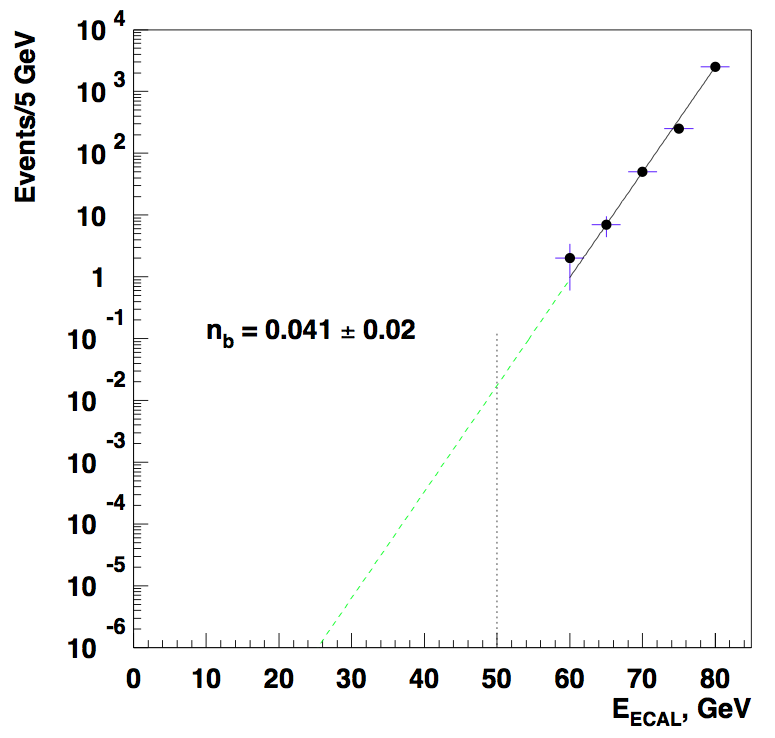}}
\vspace{-.0cm}{\caption{ Energy distribution of events in the side band C collected in 
the run II with intensity $\simeq 3.5\times 10^6~e^-$/spill and  obtained 
with pileup algorithm. 
The curve shows single exponential fit to the data, while the dashed one 
represents extrapolation to the signal region which predicts 
 $n_b = 0.041\pm 0.02$ background events.\label{fig:r2366-2436}}}
\end{figure}   

\item {\it Large transverse fluctuations.}
Another  possible source of background was caused by  the large transverse fluctuations of hadronic showers from the reaction $e Z \to eZ + \geq 2~neutrals$ 
induced by electrons in the ECAL. 
In such events all secondary long-lived neutral particles (such as neutrons and/or $K^0_L$'s) could be  produced in the target at a large angle,   the HCAL and escape the detector without depositing energy  through the lateral surface, thus resulting in the fake signal event. 
 Taking  into account results  from the previous study 
\cite{gkkk,nomad1,sngnomad},  a conservative estimate  for this background  gives  the level  $\lesssim 10^{-14}$ per incoming electron.
\end{itemize}

\par The beam backgrounds can be subdivided into two categories: upstream interactions and particle decays.  
\begin{itemize}
\item{\it Upstream interactions}. The main background sources is caused by the upstream beam  interactions with beamline  materials, such e.g. as entrance windows of the beam lines, residual gas, S1, MM1,2 etc., resulting in an admixture of   low energy  electrons with a large incoming angle  in the  beam.  Those may fake   a missing energy signal  as they still could be within acceptance of the   the spectrometer, while  some or all of the accompanying produced secondaries fall outside the acceptance of the downstream ECAL and  HCAL calorimeters. The limited detection acceptance of the secondaries  along the 
downstream beam axis enhances this background. In 2017 run a zero-angle 
 detector, as well as  the lead-glass counters are  planned to be installed 
to improve the downstream coverage and detection efficiency.  
The fraction of upstream scattered  events is estimated to be at the level 
$\lesssim 10^{-5}$ per EOT. An uncertainty arises also  from the lack of 
accurate knowledge of the dead material composition in the beam line and is  potentially the largest source of systematic uncertainty for accurate calculations of the fraction and energy distribution of these events. 
As it is not clear whether  such rare large angle scattering could be reproduced with MC simulations the amount of background
events from the beam upstream interactions was mainly estimated from the
data itself. 
\par In addition to the SRD cut which helps to reject low energy electrons and  
 $V_2$ cut, which rejects most of the charged secondaries  up to 35 cm away from the deflected beam axis, two additional cuts  were used to study possible contribution from this source. 
The first one eliminated charged secondaries  with multiple  hits  in the upstream 
tracker chambers more than expected from a single track event. 
 The second one,  used  information on  lateral reconstructed energy and time 
spread  in  the  HCAL cells from charged and neutral secondaries.  
It was used  to reject mostly events accompanied by low energy 
neutrals and charged  secondaries with typically more activity in the HCAL than expected from interaction in the ECAL target.  
In Fig.~\ref{ecvshc} the comparison of events distribution before (central panels) and after using the 
HCAL cut (right panels) is shown. One can clearly see that the amount of background events   in the vicinity of the masked signal region is substantially reduced.

Finally the background level  in the signal box  was estimated  from the extrapolation of
the number of data events observed in dedicated control regions to the signal region using the fitting procedure described below. 
Namely,  we looked at the ECAL  energy  distribution in the control region $E_{ECAL} >  50; E_{HCAL} < 1$ GeV 
for the runs I-III and estimated the 
contamination level in the signal box  by fitting 
the $E_{ECAL}$ distribution with the function   $f(E_{ECAL})= exp[p1+p2\cdot E_{ECAL}(GeV)]$, where $p1$ is a constant, and $p2$ is the slope. 
In order to  validate the exponential shape of the extrapolation function from control to signal region, dedicated
validation region $1 < E_{ECAL}  < 80$ GeV,  $E_{HCAL}  > 10  $ GeV  containing a bigger data sample of events from hadronic interactions  of the beam electrons  with nuclei of  the ECAL target   was  defined, see right panels in Fig.~\ref{ecvshc}.  The exponential shape of the distribution of energy deposited by the scattered electrons in the  ECAL  observed in the validation region was 
cross checked with MC simulations, see Fig. 5 in Ref. \cite{gkkk},  and  found to be in agreement  for  the full energy range.  Therefore, the amount of 
background in the signal region was estimated form the extrapolation procedure  assuming that the  energy distribution of  beam electrons scattered in hadronic reactions the upstream part of the beamline  has  exponential shape similar to the one observed  in hadronic interactions of beam electrons  in  the ECAL target. 
Note that background of electrons from  the upstream QED  bremsstrahlung scattering is strongly suppressed as they typically follow the beam direction after 
the scattering,  and fall outside of the detector acceptance  after being deflected  in the magnets. 

 The yield of the background events was estimated by extrapolating the fit functions from the side band C   to the signal box, see  Fig. ~\ref{ecvshc},  
 assessing the systematic uncertainties by varying the background fit functions within the corresponding errors. 
 An example of the fit extrapolation for the side C of the ECAL energy distribution is shown in Fig.~\ref{fig:r2366-2436} for the run II. 
The slopes in the exponential	fitting	of the $E_{ECAL}$ distributions for the  runs I-III  were  $0.315\pm 0.0021, ~ 0.396\pm 0.0072, 
0.49\pm0.0026$ in unit of 1/(GeV/c), and the number of $n_b$ events  
in the  signal region  were expected to be 
$0.043 \pm 0.017,~0.041 \pm 0.02,  ~0.01 \pm 0.003  $, respectively.    Possible variation 
 of the HCAL zero-energy threshold during data taking were also taken into account.  
 The  fit was also performed for both  sideband A shown in the right panel of 
Fig.~\ref{ecvshc}. Events in the region A ($E_{ECAL} < 50~{\text GeV }; E_{HCAL} > 1~{\text GeV }$) are pure neutral hadronic secondaries produced 
 by electrons in the ECAL target, while  events from the region C ($E_{ECAL}\gtrsim 50~{\text GeV }; E_{HCAL} < 1~{\text GeV }$) are likely  from the  $e^-$ 
 interactions in the upstream part of the beam line.  As a result, $\lesssim 0.001$ events in total are  expected in the signal box from the side band A for all runs, and 
was further neglected. 
 
\begin{table*}[tbh!] 
\begin{center}
\caption{Summary of estimated numbers of  background events inside the signal 
box  for   $4.3\times 10^{10}$ EOT.}\label{tab:bckg}
\vspace{0.15cm}
\begin{tabular}{lr}
\hline
\hline
Background source& Estimated number of events, $n_b$\\
\hline
hermeticity: punchthrough $\g$'s, cracks, holes & $<0.001$\\
loss of hadrons from $e^-Z\to e^- + hadrons$ & $<0.001$\\
loss of muons from $\emu$&$0.005\pm 0.001$\\
$\mu\to e \nu \nu$, $\pi,~K \to e \nu$, $K_{e3}$ decays  & $ 0.02\pm 0.004$ \\
$e^-$ interactions in the beam line materials&$0.09\pm 0.03$\\
$\mu, \pi, K$ interactions in the target & $0.008\pm 0.002$ \\
accidental SR tag and $e^-$ from $\mu,\pi, K$ decays & $<0.001$ \\
\hline 
Total $n_b$   &    $0.12\pm 0.04$\\
\hline
\hline 
\end{tabular}
\end{center}
\end{table*}
 Finally, the contamination of backgrounds to the signal region due to beam interactions was estimated to be  $0.09\pm 0.03$ events.
The uncertainty in the background estimate due to upstream scattered events 
was dominated by the systematic uncertainty of the upstream veto V1 and tracker 
efficiency, precise knowledge of the material  in the beam line, and 
statistics of the data samples. This systematic uncertainty was estimated 
by performing measurements on several samples of upstream scattered events 
tagged by a signature of scattering in the HCAL.     
The uncertainties in
this background estimate are evaluated by considering differences in the estimates of the event number   by varying the electron identification probabilities and
changing the parameters of the extrapolation functions.

\item {\it Particle decays}. 
 Other backgrounds were expected from the decays of  $\mu,~\pi,~K \to e + É$  in flight in the beamline accompanied by emission of an energetic neutrino.
These  backgrounds were highly suppressed by requiring the presence of the 
SRD tag. However, there might be cases  when,  e.g.  a pion could  knock electrons off the downstream window of the vacuum vessel, which hit the SRD creating a fake tag for a 100 GeV $e^-$. The pion could then decay into $e\nu$ in the 
upstream ECAL region thus producing the fake signal. 
\par The main  background source in this category was $K_{e3}$ decays where the electron overlapped 
with photons from $\pi^0$ decay thus producing a single-like e-m shower in the 
ECAL. In addition to the SRD cut and the probability of  decay in the downstream 
part of the setup, this process was further suppressed  by requiring 
shower energy to be $< 50$ GeV and the incoming track azimuthal angle to be 
below 5 mrad. The transverse and longitudinal 
shower shape at the ECAL was also used to distinguish the single electron 
shower from the overlapped one. 

Similar background was caused by  a random superposition of  uncorrelated low-energy,  50 - 70 GeV,  electron  from the low-energy beam tail and 100 GeV beam 
$\mu,~\pi, ~ K$  occurring during the detector gate-time. 
The electron could emit the amount of SR energy above the threshold which is detected in the SRD as a tag of  100 GeV $e^-$ and then is deflected by the spectrometer magnets out of the detector's acceptance angle. While the  accompanying  mistakenly tagged $\mu,\pi$ or $K$ could either decays in-flight in front of the ECAL into the $e^- + X$ state with the decay electron energy less then the beam energy, or 
could also interact in the target producing an e-m like cluster below 50 GeV though the $\mu Z \to \mu Z \g$ or $\pi, K$ charge-exchange reactions, 
accompanied by the poorly detected scattered $\mu$, or secondary hadrons, thus resulting in both cases to  the signal signature of Eq.(\ref{sign}).  
 These  background components were  simulated with a statistics higher or comparable to the  number of events expected from  the data and was  found to be small. 
\end{itemize}
The remaining physical backgrounds were 
\begin{itemize}
\item {\it Dimuon, $\tau$, charm decays}. 
 The process \eqref{dimu} could mimic the signal either i) due to muons decay  in flight inside  the ECAL target  into $e\nu \nu$ state, or ii) if the muons escape  detection  in the $V_2$ and HCAL modules due to  fluctuations of the energy   (number of photoelectrons) deposited  in these detectors.
   In the case i)  the relatively long muon lifetime results  in a small probability to decay inside the ECAL. 
 For the case ii) the background is suppressed by the high-efficiency veto system $V_2 $+HCAL. 
 The $V_2$ was  a  $\sim 4$  cm thick  high-sensitivity scintillator  array whose inefficiency 
for a single muon  detection  was estimated to be  $\lesssim 10^{-4}$. Therefore, the level of  dimuon  background is expected to be  $< 10^{-13}$ per EOT. 
The fake signal  could also arise from the reactions of $\tau$, e.g., $eZ \to  eZ \tau^+ \tau^-; ~ \tau \to e \nu \nu $,  or charm, e.g., 
$eZ \to  eZ +D_s+anything; ~ D_S\to e+\nu+anything  $, production and their subsequent 
prompt decays into an electron accompanied by emission of neutrinos.  The estimate show that these backgrounds are also 
expected to be negligible. 

\item Finally,  the electroproduction of a neutrino pair 
$eZ \to  eZ \nu \overline{\nu}$ resulting in the invisible final state accompanied by energy deposition in the ECAL1 from the recoil electron  can occur. 
An  estimate showed that the ratio
of the cross sections for this  reaction  to the bremsstrahlung cross section 
is well below $10^{-13}$ \cite{Gninenko:2013rka}. 
\end{itemize}
\par In Table~\ref{tab:bckg} the contributions from all background processes 
estimated by using the MC simulations, exept for those from beam interactions in the upstream part of the setup,  are summarized. 
The final number of background events  estimated from the combined MC and data 
events  is  $n_{b}=0.12 \pm 0.04$ events for $4.3\times 10^{10}$ EOT.
 The estimated uncertainty of about 30\% was due mostly to the uncertainty 
in background level from upstream beam interactions. It also  includes the uncertainties in the amount of passive material 
for  $e^-$ interactions,  in the  cross sections of the hadron  charge-exchange reactions on lead (30\%), and  systematic errors
related to the extrapolation procedure.
The total systematic uncertainty  was calculated by adding all errors in quadrature.\ 
\begin{figure}[tbh!!]
\begin{center}
\vspace{-0cm}{\includegraphics[width=.5\textwidth, height=.4\textwidth]{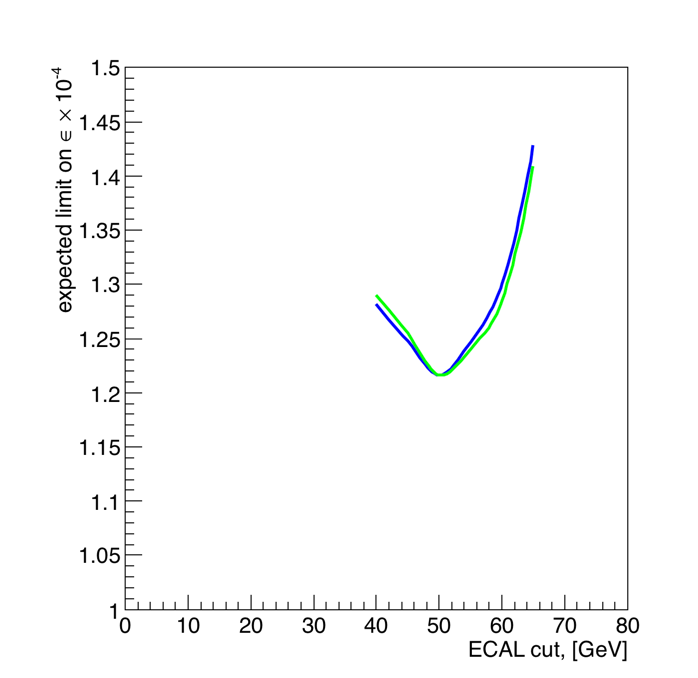}}
\caption {The sensitivity,   defined as an average expected limit,  as a function of the ECAL energy cut for the case of  the $A'$ detection
with the mass $\ma \simeq 20 $ (blue) and 2 (green) MeV.  }
\label{opt}
\end{center}
\end{figure}
\section{Results and calculation of limits} 
\label{sec:results}
\par In the final statistical analysis the three runs I-III were analysed simultaneously using the multi-bin limit setting technique.
The corresponding code is based on the RooStats package \cite{root}. First of all,
the above obtained  background estimates, efficiencies,  and their  corrections and uncertainties were used 
to optimize more accurately the main cut defining the signal box by comparing
sensitivities,  defined as an average expected limit calculated using the profile likelihood method, 
with uncertainties used as nuisance parameters. Log-normal distribution was assumed for the nuisance parameters \cite{Gross:2007zz}.
The most important inputs for this optimization were the expected values from the background extrapolation into the signal box for the data 
samples of the runs I,II,III. The uncertainties for background prediction were estimated by varying
the extrapolation functions, as previously discussed.
An example  of the optimization curves obtained for the $\ma = 2 $ and 20  MeV is shown in Fig.~\ref{opt}. 
It was  found that the  optimal  cut value  depends very weakly on the $A'$ mass choice and can be safely set to $E_{ECAL} < 50 $ GeV
for the whole mass range.  

\par Overall optimization and improvement of the signal selection and background rejection 
criteria resulted in roughly more than a factor 10 reduction of the expected 
backgrounds per EOT and an increase of a factor 2 in the efficiency of 
$\ainv$ decay at higher beam rate for the run III compared to those obtained in the analysis reported in Ref.\cite{na64prl}.
For the full 2016  exposure, the  estimate of the number of background events expected from 
the sources discussed above per $10^{10}$ EOT was $n_b=0.03$,
while for the run of Ref.\cite{na64prl} it was $n_b=0.5$.   

\begin{figure}[tbh]
\begin{center}
\includegraphics[width=0.5\textwidth]{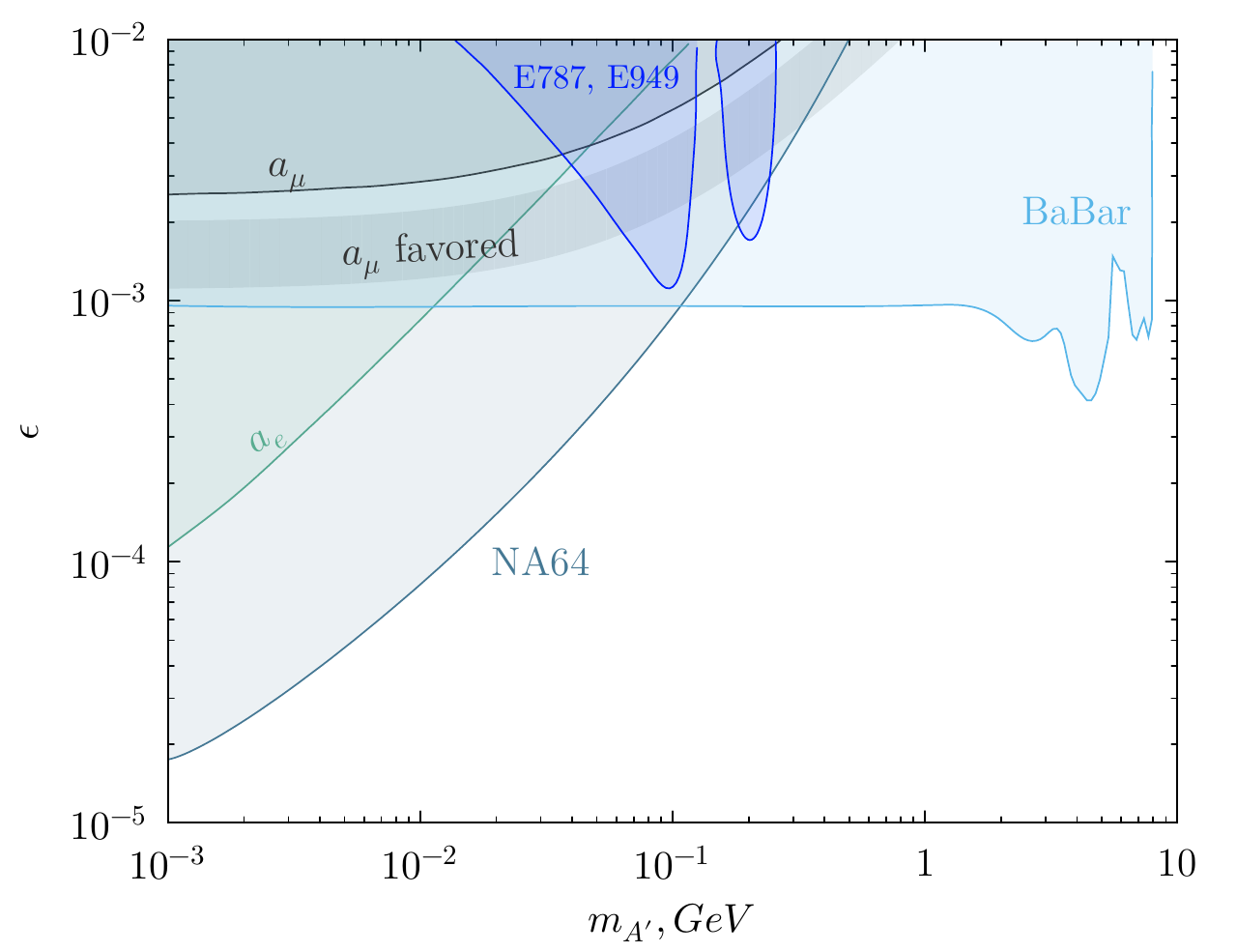}
\caption {The NA64 90\% C.L. exclusion region in the ($m_{A'}, \epsilon$) 
plane.   Constraints from the BaBar \cite{babarg-2},  
 E787 and  E949  experiments \cite{hd,Essig:2013vha}, as well as the muon  $\alpha_\mu$ favored area 
 are also shown. Here, $\alpha_\mu =\frac{g_\mu-2}{2}$.
 For more limits obtained from indirect searches and 
 planned measurements see e.g. Ref. \cite{report1,report2}.
  \label{exclinv}}
\end{center}
\end{figure} 
\begin{table}[tbh!!]
\begin{center}
\begin{tabular}{|c|c|c|}
\hline
\hline
$\ma$, MeV &  90\% C.L. upper limit       &    90\% C.L. upper limit  \\
& on $\epsilon,~10^{-4}$ ,  no $k$-factors & on $\epsilon,~10^{-4} $with  $k$-factors \\
\hline
1.1 & 0.22 & 0.19\\
2 &0.23& 0.24\\
5&0.43& 0.49\\
16.7&1.25&1.33\\
20&1.29& 1.6\\
100 &5.5&8.2\\
200 &13.0&22.6\\
500 &38.7&97.8\\
950 &94.20& 362.0\\
\hline
\hline
\end{tabular}
\end{center}
\caption{ Comparison of upper bounds on mixing $\epsilon$ at 90
      \% CL obtained with    WW and  ETL calculations for the 
       Pb-Sc ECAL target  for  $E_{miss} > 0.5 E_0$ at $E_0=100$ GeV.   }
 \label{tab:epscomp}
\end{table}
\begin{figure*}[tbh!]
\includegraphics[width=.45\textwidth]{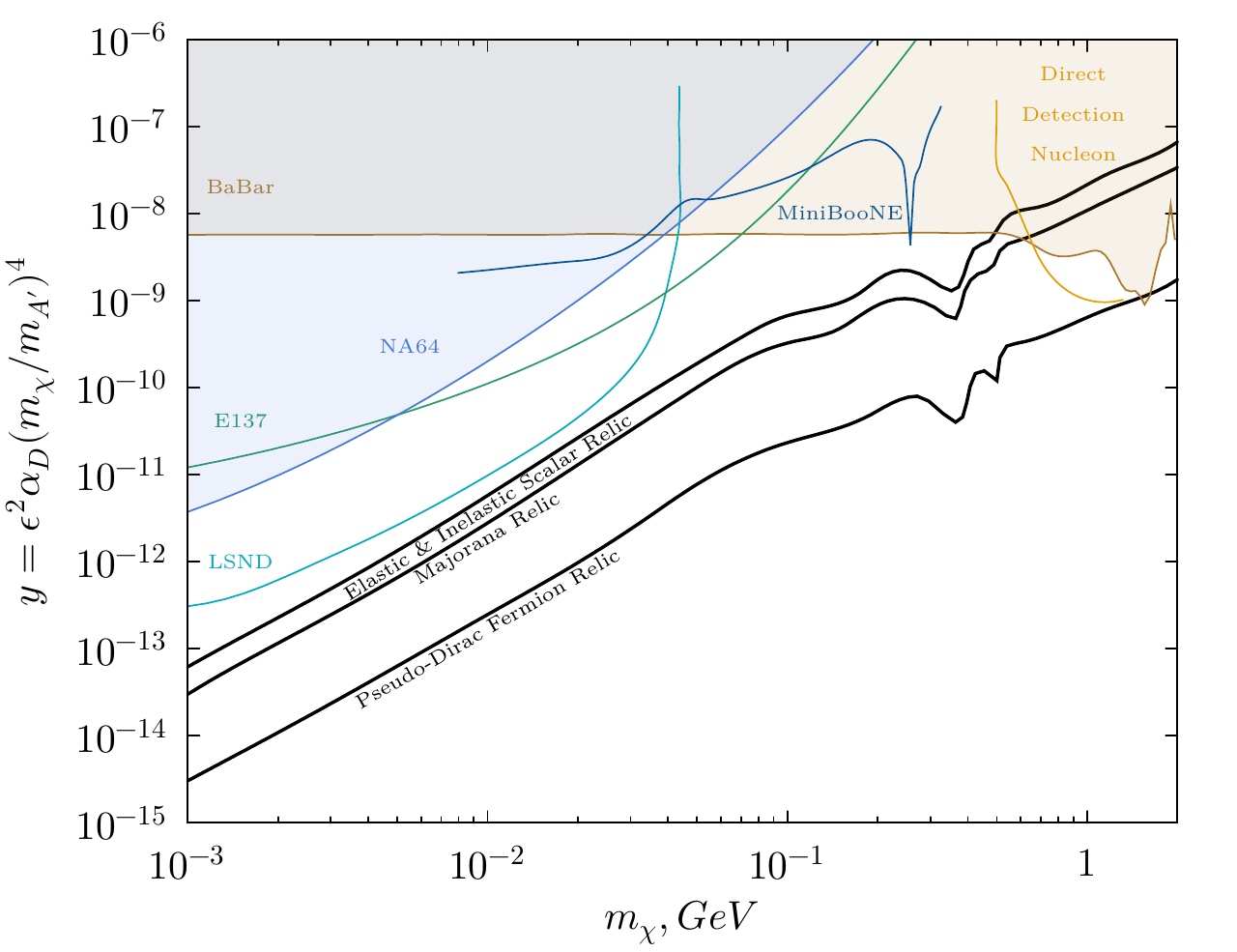}
\includegraphics[width=.45\textwidth]{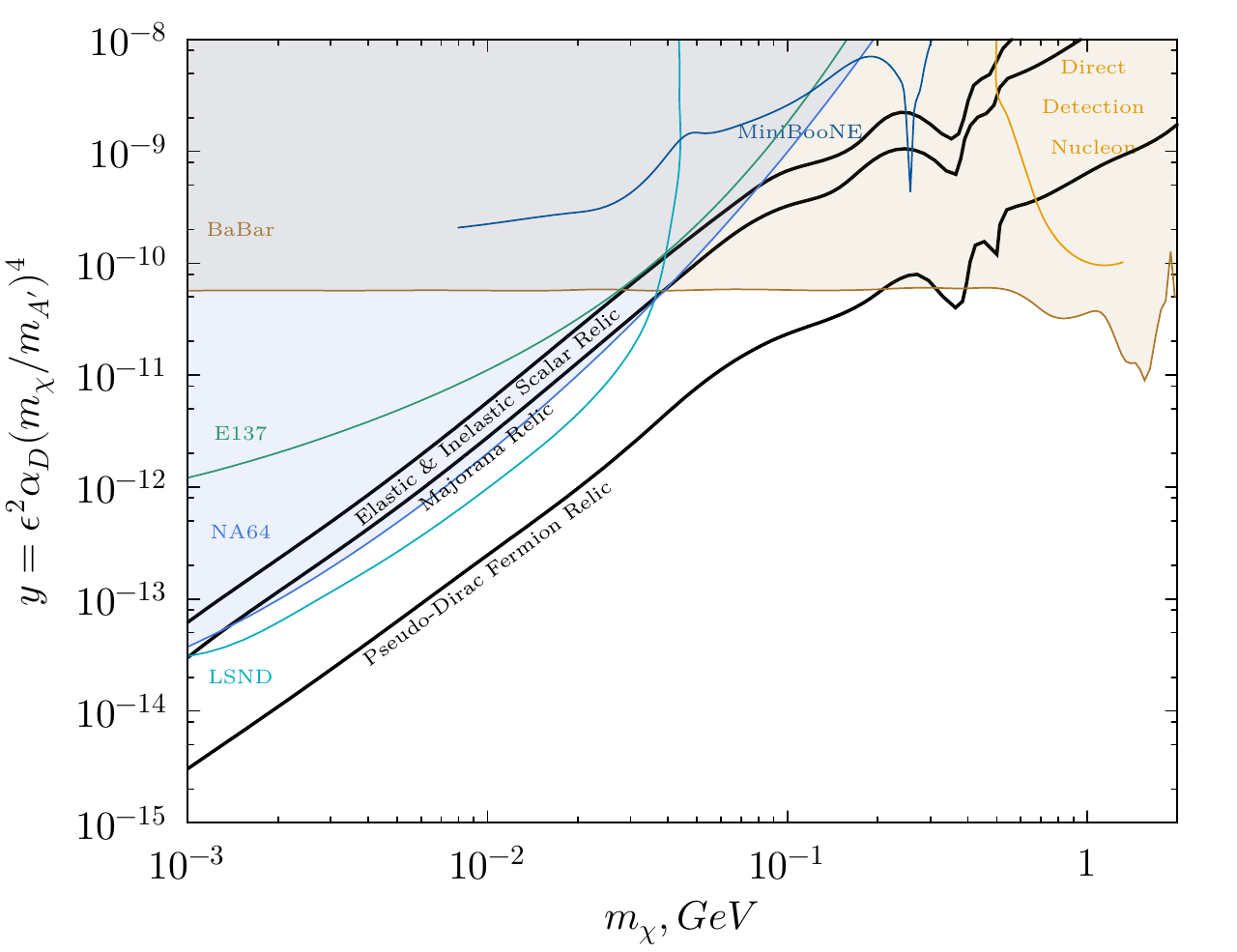}
\caption{The NA64 limits  in  the (y;$m_{\chi}$) plane obtained for $\alpha_D=0.5$ (left panel) and $\alpha_D=0.005$ (right panel) from the full 2016  data set shown in comparison with limits  obtained in Refs.\cite{report1,Izaguirre:2014bca,Iza2015,Iza2017}  from  the results of the 
LSND~\cite{deNiverville:2011it,Batell:2009di},   E137 \cite{e137th}, BaBar \cite{babarg-2}, MiniBooNE \cite{minib2017} and direct detection \cite{mardon}experiments.
The favoured parameters to account for the observed relic DM density for the  scalar, pseudo-Dirac  and Majorana  type of light thermal DM are shown as the lowest solid line.}
\label{yvsm}
\end{figure*}   
 After determining and optimizing  all the selection criteria and estimating background levels, we examined the events in the signal box and
found no candidates, as shown  in  Fig.~\ref{ecvshc}. We proceeded then with the calculation of the upper limits on the $A'$ production.
The combined 90\%  confidence level (C.L.)  upper limits for the corresponding
mixing strength  $\epsilon$  were determined from the 90\% C.L.  upper limit for the 
expected number of signal events, $N_{A'}^{90\%}$  by using the modified frequentist approach for confidence levels (C.L.), taking the profile
likelihood as a test statistic in the asymptotic approximation \cite{junk,limit,Read:2002hq}. The total number of expected signal events in the
signal box was the sum of expected events from the three runs:
\begin{equation}
\Na = \sum_{i=1}^{3} N_{A'}^i = \sum_{i=1}^{3} n_{EOT}^i  \epsilon_{tot}^i n_{A'}^i(\epsilon,\ma, \Delta E_{e})
\label{nev}
\end{equation}
where $\epsilon_{tot}^i$ is the signal efficiency in the run i given by Eq.(\ref{eff}), and the $n_{A'}^i(\epsilon,\ma, \Delta E_{A'})$ value is the 
signal yield per EOT generated by a single 100 GeV electron in the ECAL target in the energy  range $\Delta E_{e} $. 
Each $i$-th entry in this sum was calculated by simulating the signal events for  corresponding beam running conditions and  processing them through 
the reconstruction program  with the same selection criteria and  efficiency corrections as for the  data sample from  the run-i.
The expected backgrounds and estimated systematic errors   were also taking into account in the limits calculation.
The combined 90\% C.L. exclusion limits on the mixing
strength as a function of the $A'$ mass can be seen  in Fig.~\ref{exclinv}.
In  Table \ref{tab:epscomp} the  limits obtained with the  ETL  and WW calculations for different $\ma$ values are also shown for comparison.
One can see that the corrections are mostly relevant in the higher mass region $\ma \gtrsim 100$ MeV.  
The derived bounds are the best for the mass range
$0.001\lesssim \ma \lesssim 0.1 $ GeV obtained from direct searches of  $\ainv$ decays \cite{pdg}.
\par The limits were also calculated with  a  simplified method by  merging all three runs into a single run
 as described previously by Eq.(\ref{nev}).
 The  total error for the each $N_{A'}^i$ value includes the corresponding  systematic uncertainties  calculated by adding contributions from all 
 sources  in quadrature, see Sec.\ref{sec:syst}. 
 In accordance with the  $CL_s$ method \cite{Read:2002hq}, for zero number of observed events the 
90\% C.L. upper limit for the  number of signal events  is $N_{A'}^{90\%}(m_{A'})=2.3$.
Taking this and  Eq.(\ref{nev}) into account  and using the relation
 $ N_{A'}(m_{A'}) < N_{A'}^{90\%}(m_{A'}) $  resulted  in  the $90\%$ C.L. limits  in the ($m_{A'};\epsilon $) plane which 
 agreed with the one shown in Fig.~\ref{exclinv}  within a few \%. 
\section{Constraints on light thermal Dark Matter}
\label{sec:ltdm}
As discussed previously, the possibility of the existence of  light 
thermal Dark Matter (LTDM)  
has been the subject of intense theoretical activity over the past several 
years  \cite{report1,report2}, see also \cite{ltdm1,ltdm2}. 
The LTDM models  can be classified by the 
spins and masses of the DM particles  and mediators.  The scalar dark  matter  mediator models are severely restricted or even excluded  by non-observation  of rare  B-meson decays \cite{report2,pdg}, so we consider here only 
the case of a vector mediator.
\begin{figure*}[tbh!]
\includegraphics[width=0.45\textwidth,height=0.4\textwidth]{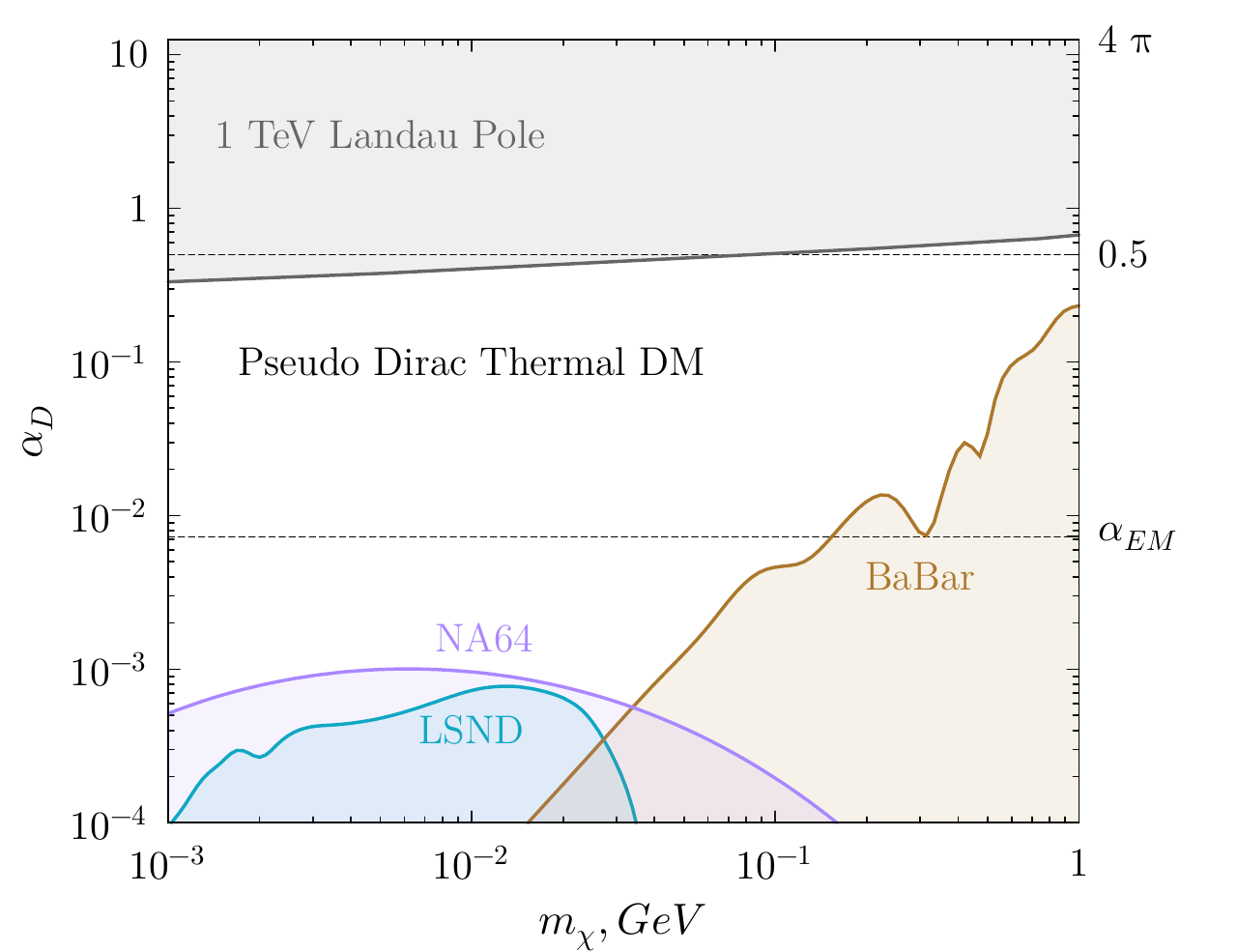}
\hspace{-0.cm}{\includegraphics[width=0.45\textwidth,height=0.4\textwidth]{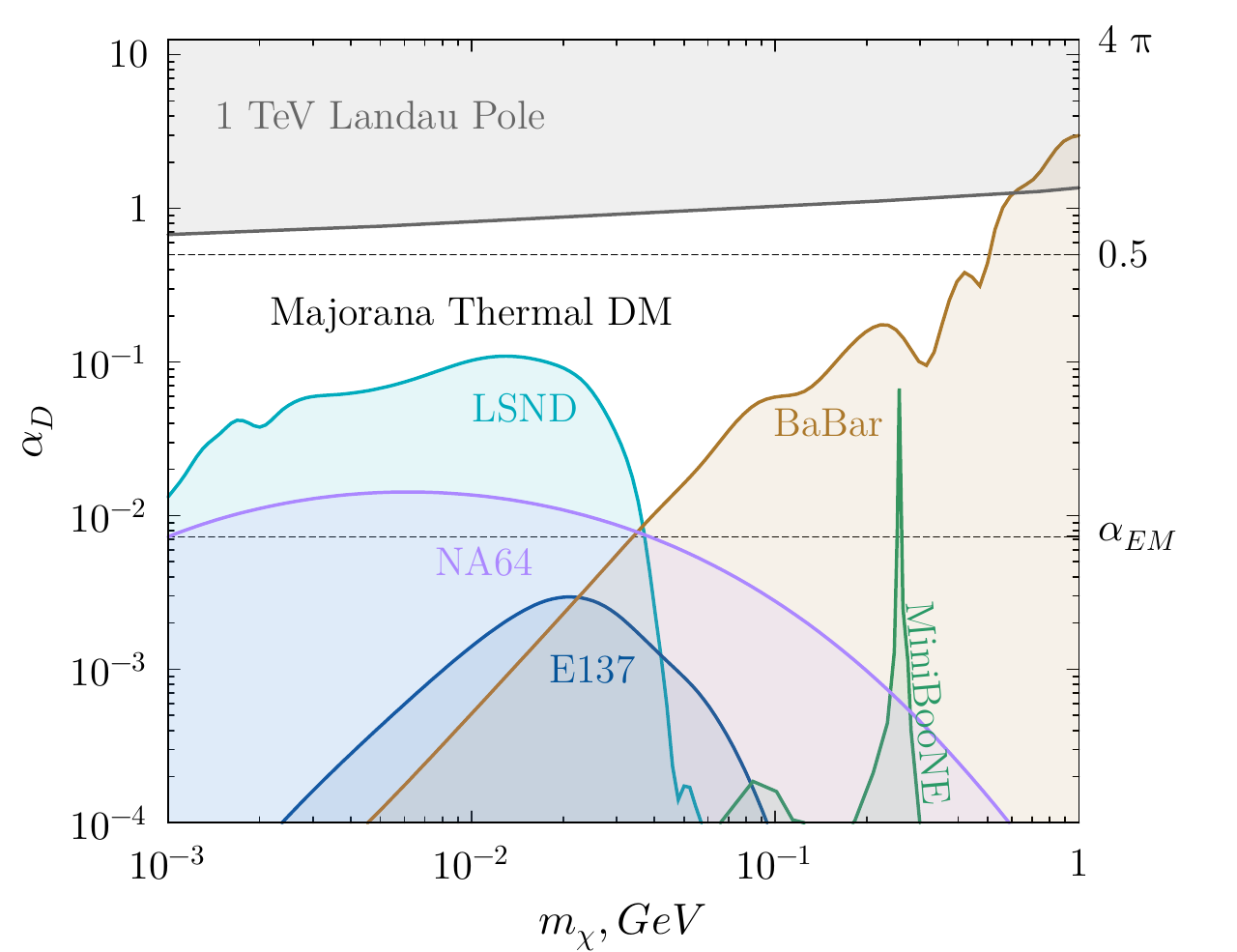}}
\caption{The NA64 constraints in the ($\alpha_D$;$m_{A'}$) plane  on  the pseudo-Dirac (the left panel) 
and Majorana ( right panel) type light thermal DM shown in comparison with bounds obtained in Ref. \cite{report2} from  the results of the 
LSND~\cite{deNiverville:2011it,Batell:2009di},   E137 \cite{e137th}, BaBar \cite{babarg-2}  and MiniBooNE \cite{minib2017} experiments.}
\label{alphavsm}
\end{figure*}   
	 As was  discussed in Sec.\ref{sec:intro}, the most popular vector mediator model is 
the one with additional massive dark  photon $A'$ which couples 
with DM particles via interaction $L  = e_DA'_{\mu}J^{\mu}_\chi$. The currents 
$J^{\mu}_{\chi} = \bar{\psi}_{\chi}\gamma^{\mu}\psi_{\chi}$ and $ J^{\mu}_{\chi} = i(\phi^{+}_{\chi}\partial^{\mu}\phi_{\chi}-
\phi_{\chi}\partial^{\mu}\phi^+_{\chi})$ for spin $1/2$ and $0$, respectively. Here, $\chi$  denotes both, either scalar or fermion LTDM particle. 
 As discussed in Sec.\ref{sec:intro}, 
the $\g - A'$ mixing  leads to nonzero interaction of dark photon $A'_{\mu}$ with the  electrically charged  SM particles with the charges 
$e' =  \epsilon e$.  As a  result of the mixing the cross-section of DM 
particle annihilation into SM particles, which determines the relic 
DM density, is proportional to $\epsilon^2$. Hence using  
constraints on the cross section of  the DM annihilation freeze out
 (resulting in  Eq.(\ref{alphad})), and obtained limits on mixing 
strength of Fig.~\ref{exclinv}, one  can derive constraints 
 in the ($y$;$m_{\chi}$) plane,  which 
 can also be used to restrict  models predicting  existence of LTDM
for the masses  $m_{\chi} \lesssim 1$~GeV. 
\par These limits obtained  from the full
data sample of the 2016 run  are shown in  the left panel of Fig.~\ref{yvsm} together with 
the favoured parameters for scalar,  pseudo-Dirac (with a small splitting)  and Majorana  scenario  of  LTDM   taking into account  the observed relic DM density \cite{report2}. The limits  are calculated by using  Eq.(\ref{y})
   under the conventional assumption $\alpha_D= 0.5$, and  $m_{A'}=3 m_{\chi}$, here $m_\chi $ stands for the LTDM 
particle's masses,  either scalars or fermions. 
The plot shows also the comparison of our results with limits  from other experiments.  Note, that some of these limits were 
obtained by using WW approximation for the cross section calculation and therefore might require revision.
The choice of $\alpha_D = 0.5$ is
compatible with the bounds derived in Ref. \cite{davou} based on
the running of the dark gauge coupling. However, it  should be 
 noted  that differently form the results of  beam dump experiments, such as  LSND~\cite{deNiverville:2011it,Batell:2009di},   E137 \cite{e137th}, MiniBooNE \cite{minib2017},    the $\chi$-yield in our case scales as $\epsilon^2$, not  as  $\epsilon^4 \alpha_D$. Therefore, 
for sufficiently small values of $\alpha_D$ our  limits  will be much stronger. This is 
illustrated in the right panel of  Fig.~\ref{yvsm},  where the NA64 limits and bounds from other experiments are shown for  
 $\alpha_D = 0.005$. One can see,  that  for this, or smaller, values of   $\alpha_D$,  the direct search for the $\ainv$ decay in NA64 
  excludes  model of scalar  and Majorana DM production via vector mediator   for the  remaining   mass region  $m_\chi \lesssim 0.05$ GeV. 
  While being combined with the BaBar limit \cite{babarg-2}, the result  excludes the model for the  entire mass region $m_\chi \lesssim 1$ GeV. 
\par The experimental upper bounds on $\epsilon$ also 
 allow to obtain lower bounds on coupling constant $\alpha_D$ 
which are shown in Fig.~\ref{alphavsm}  in the ($\alpha_D$;$m_{\chi}$) plane.
 For the case of pseudo-Dirac fermions and small splitting,   the limits  in the left panel of Fig.~\ref{alphavsm} were calculated by taking 
 the value $f=0.25 $ in Eq.(\ref{alphad}).  For the mass range $m_{\chi} \lesssim 0.05$ GeV 
 the obtained bounds are more stringent than the limits obtained from the results of  LSND~\cite{deNiverville:2011it,Batell:2009di} and  E137 \cite{e137th}.  
 The limits for the Majorana case   shown in the right panel of Fig.~\ref{alphavsm} were calculated by setting $f=3$.  To cross check our calculations,  we also  derived  limits on  $\alpha_D$   by using  BaBar bounds on $\epsilon$ \cite{babarg-2}, see  Fig.~\ref{exclinv},   Eq.(\ref{alphad}) and the previous $f$ values for the pseudo-Dirac and Majorana cases. 
The obtained  BaBar limits were found to be in  good agreement with those 
shown in Fig.~\ref{alphavsm} for the mass region $m_\chi \lesssim  0.1$ GeV.  
Note,  that  new constraints for the large pseudo-Dirac fermion splitting   
 can also be derived. They will be more stringent than for the case of the small splitting 
 and similar to the one obtained for the Majorana case. 

\section{Conclusion}
\label{sec:conclusion}

From the analysis of the full 2016 data sample, we found no evidence for the existence of  dark photon with  the mass  in the 
range $\lesssim 1 $ GeV  which  mixes  with the 
ordinary photon  and decays dominantly invisibly into light DM 
particles  $A'\to \chi \overline{\chi}$.\  
New limits on the mixing strength were derived  by taking into account  the $A'$ production cross sections calculated at the exact tree-level  in  Ref.\cite{kk}  
without using  Weizs\"{a}cker-Williams approximation. These cross sections, implemented  into NA64 simulation package,  
 were cross checked with the one calculated by Liu et el. 
following their approach reported in Ref.\cite{Liu:2016mqv,Liu:2017htz}  and was found to be in agreement. 
Good agreement between the data and MC  for the  rare QED dimuon  production in the reaction \eqref{dimu}   was also  found. This process was used as a robust  benchmark  for the signal event simulation and analysis. 
 For the mass range $\ma\lesssim 0.1$ GeV the most stringent  upper limit on  the  
mixing strength, $\epsilon$ were   obtained.\
Using conventional choices  for  the DM parameters we also set  the 90\% C.L. limits
on the value $y$,   representing the dimensionless DM annihilation cross section parameter, for the pseudo-Dirac and Majorana DM in the  $\chi$ mass region    $0.001 < m_\chi < 0.1$ GeV. With these
DM parameter combinations,  our  result  has expanded
the search for DM to $y$ values an  order of magnitude
smaller than  MiniBooNE  DM experiment \cite{minib2017}. 

\par For  the vector portal DM model and the chosen parameter constraints, 
 the  obtained lower limits  $\alpha_D \gtrsim 10^{-3} $ for pseudo-Dirac Dark Matter in the mass region 
 $ m_\chi \lesssim  0.05 $ GeV are  more stringent than the bounds from beam dump experiments.
 For  values $\alpha_D \lesssim0.005$  the  combined results from direct searches of the $\ainv$ decay in NA64 and BaBar experiments   
 exclude   model for scalar   and Majorana DM production via vector portal  for  the  mass region $m_\chi \lesssim 1$ GeV. 
The obtained results  are used to constrain an interpretation of the $A'$ as a mediator of light thermal DM production. 
The  remaining windows of  parameter space  between our result and the Landau pole bounds obtained from arguments 
 based on the running of the dark gauge coupling \cite{davou}  for  these  scenarios  can be covered  through future  searches with the NA64 experiment. 
 This   would require an improved  sensitivity of about  two  orders of magnitude which is feasible. 

{\large \bf Acknowledgements}
 We gratefully acknowledge the support of the CERN management and staff 
and the technical staffs of the participating institutions for their vital contributions. 
 This work was supported by the HISKP, University of Bonn (Germany), JINR (Dubna), MON, RAS and  RF program ``Nauka''  (Contract No. 0.1764.GZB.2017) (Russia),  ETH Zurich and SNSF Grant No. 169133 (Switzerland), and  grants FONDECYT 1140471, 1150792, and 3170852, Ring ACT1406 and Basal FB0821 CONICYT (Chile).
Part of the work on MC simulations was supported by the RSF grant 14-22-00161.  
 We thank Gerhard Mallot for  useful suggestions,  Maxim Pospelov for useful communications and help provided for Sec. IX, and   Didier  Cotte, Michael Jeckel, Vladimir Karjavin, Christophe Menezes Pires,  and  Victor Savrin for their help. 
 We thank COMPASS DAQ group and the Institute for Hadronic Structure and Fundamental Symmetries of TU Munich for the technical support.

\end{document}

%% file: author_list.tex
\affiliation{\it Universit\"at Bonn, Helmholtz-Institut f\"ur Strahlen-und Kernphysik, 53115 Bonn, Germany} 
\affiliation{\it  Joint Institute for Nuclear Research, 141980 Dubna, Russia}
\affiliation{\it CERN, European Organization for Nuclear Research, CH-1211 Geneva, Switzerland}
\affiliation{\it Institute for Nuclear Research, 117312 Moscow, Russia}
\affiliation{\it P.N. Lebedev Physics Institute, Moscow, Russia, 119 991 Moscow, Russia}
\affiliation{\it Skobeltsyn Institute of Nuclear Physics, Lomonosov Moscow State University, 119991 Moscow, Russia}
\affiliation{\it Technische Universit\"at M\"unchen, Physik  Department, 85748 Garching, Germany}
\affiliation{\it Physics Department, University of Patras, 265 04 Patras, Greece} 
\affiliation{\it State Scientific Center of the Russian Federation Institute for High Energy Physics of National Research Center 'Kurchatov Institute' (IHEP), 142281 Protvino, Russia}
\affiliation{\it Tomsk Polytechnic University, 634050 Tomsk, Russia}
\affiliation{\it Universidad T\'{e}cnica Federico Santa Mar\'{i}a, 2390123 Valpara\'{i}so, Chile}
\affiliation{\it ETH Z\"urich, Institute for Particle Physics, CH-8093 Z\"urich, Switzerland}
\author{D.~Banerjee}\affiliation{\it ETH Z\"urich, Institute for Particle Physics, CH-8093 Z\"urich, Switzerland}
\author{V.~E.~Burtsev}\affiliation{\it Tomsk Polytechnic University, 634050 Tomsk, Russia}
\author{A.~G.~Chumakov}\affiliation{\it Tomsk Polytechnic University, 634050 Tomsk, Russia}
\author{D.~Cooke}\affiliation{\it ETH Z\"urich, Institute for Particle Physics, CH-8093 Z\"urich, Switzerland}
\author{P.~Crivelli}\affiliation{\it ETH Z\"urich, Institute for Particle Physics, CH-8093 Z\"urich, Switzerland}
\author{E.~Depero}\affiliation{\it ETH Z\"urich, Institute for Particle Physics, CH-8093 Z\"urich, Switzerland}
\author{A.~V.~Dermenev}\affiliation{\it Institute for Nuclear Research, 117312 Moscow, Russia}
\author{S.~V.~Donskov}\affiliation{\it State Scientific Center of the Russian Federation Institute for High Energy Physics of National Research Center 'Kurchatov Institute' (IHEP), 142281 Protvino, Russia}
\author{F.~Dubinin}\affiliation{\it P.N. Lebedev Physics Institute, Moscow, Russia, 119 991 Moscow, Russia}
\author{R.~R.~Dusaev}\affiliation{\it Tomsk Polytechnic University, 634050 Tomsk, Russia}
\author{S.~Emmenegger}\affiliation{\it ETH Z\"urich, Institute for Particle Physics, CH-8093 Z\"urich, Switzerland}
\author{A.~Fabich}\affiliation{\it CERN, European Organization for Nuclear Research, CH-1211 Geneva, Switzerland}
\author{V.~N.~Frolov}\affiliation{\it  Joint Institute for Nuclear Research, 141980 Dubna, Russia}
\author{A.~Gardikiotis}\affiliation{\it Physics Department, University of Patras, Patras, Greece}
\author{S.~G.~Gerassimov }\affiliation{\it P.N. Lebedev Physics Institute, Moscow, Russia, 119 991 Moscow, Russia}\affiliation{\it Technische Universit\"at
M\"unchen, Physik Dept., 85748 Garching, Germany}
\author{S.~N.~Gninenko\footnote{Corresponding author, Sergei.Gninenko@cern.ch}}\affiliation{\it Institute for Nuclear Research, 117312 Moscow, Russia}
\author{M.~H\"osgen}\affiliation{\it Universit\"at Bonn, Helmholtz-Institut f\"ur Strahlen-und Kernphysik, 53115 Bonn, Germany}
\author{A.~E.~Karneyeu}\affiliation{\it Institute for Nuclear Research, 117312 Moscow, Russia}
\author{B.~Ketzer}\affiliation{\it Universit\"at Bonn, Helmholtz-Institut f\"ur Strahlen-und Kernphysik, 53115 Bonn, Germany}
\author{D.~V.~Kirpichnikov}\affiliation{\it Institute for Nuclear Research, 117312 Moscow, Russia}
\author{M.~M.~Kirsanov}\affiliation{\it Institute for Nuclear Research, 117312 Moscow, Russia}
\author{I.~V.~Konorov}\affiliation{\it P.N. Lebedev Physics Institute, Moscow, Russia, 119 991 Moscow, Russia} \affiliation{\it Technische Universit\"at
M\"unchen, Physik Dept., 85748 Garching, Germany}
\author{S.~G.~Kovalenko}\affiliation{\it Universidad T\'{e}cnica Federico Santa Mar\'{i}a, 2390123 Valpara\'{i}so, Chile}
\author{V.~A.~Kramarenko}\affiliation{\it  Joint Institute for Nuclear Research, 141980 Dubna, Russia}\affiliation{\it Skobeltsyn Institute of Nuclear Physics, Lomonosov Moscow State University, Moscow, Russia}
\author{L.~V.~Kravchuk}\affiliation{\it Institute for Nuclear Research, 117312 Moscow, Russia}
\author{ N.~V.~Krasnikov}\affiliation{\it Institute for Nuclear Research, 117312 Moscow, Russia}
\author{S.~V.~Kuleshov}\affiliation{\it Universidad T\'{e}cnica Federico Santa Mar\'{i}a, 2390123 Valpara\'{i}so, Chile}
\author{V.~E.~Lyubovitskij}\affiliation{\it Tomsk Polytechnic University, 634050 Tomsk, Russia}\affiliation{\it Universidad T\'{e}cnica Federico Santa Mar\'{i}a, 2390123 Valpara\'{i}so, Chile}
\author{V.~Lysan}\affiliation{\it  Joint Institute for Nuclear Research, 141980 Dubna, Russia}
\author{V.~A.~Matveev}\affiliation{\it  Joint Institute for Nuclear Research, 141980 Dubna, Russia}
\author{Yu.~V.~Mikhailov}\affiliation{\it State Scientific Center of the Russian Federation Institute for High Energy Physics of National Research Center 'Kurchatov Institute' (IHEP), 142281 Protvino, Russia}
\author{D.~V.~Peshekhonov}\affiliation{\it  Joint Institute for Nuclear Research, 141980 Dubna, Russia}
\author{V.~A.~Polyakov}\affiliation{\it State Scientific Center of the Russian Federation Institute for High Energy Physics of National Research Center 'Kurchatov Institute' (IHEP), 142281 Protvino, Russia}
\author{B.~Radics}\affiliation{\it ETH Z\"urich, Institute for Particle Physics, CH-8093 Z\"urich, Switzerland}
\author{R.~Rojas}\affiliation{\it Universidad T\'{e}cnica Federico Santa Mar\'{i}a, 2390123 Valpara\'{i}so, Chile}
\author{A.~Rubbia}\affiliation{\it ETH Z\"urich, Institute for Particle Physics, CH-8093 Z\"urich, Switzerland}
\author{V.~D.~Samoylenko}\affiliation{\it State Scientific Center of the Russian Federation Institute for High Energy Physics of National Research Center 'Kurchatov Institute' (IHEP), 142281 Protvino, Russia}
\author{V.~O.~Tikhomirov}\affiliation{\it P.N. Lebedev Physics Institute, Moscow, Russia, 119 991 Moscow, Russia}
\author{D.~A.~Tlisov}\affiliation{\it Institute for Nuclear Research, 117312 Moscow, Russia} 
\author{A.~N.~Toropin}\affiliation{\it Institute for Nuclear Research, 117312 Moscow, Russia}
\author{A.~Yu.~Trifonov}\affiliation{\it Tomsk Polytechnic University, 634050 Tomsk, Russia}
\author{B.~I.~Vasilishin}\affiliation{\it Tomsk Polytechnic University, 634050 Tomsk, Russia}
\author{G.~Vasquez Arenas}\affiliation{\it Universidad T\'{e}cnica Federico Santa Mar\'{i}a, 2390123 Valpara\'{i}so, Chile}
\author{P.~Ulloa}\affiliation{\it Universidad T\'{e}cnica Federico Santa Mar\'{i}a, 2390123 Valpara\'{i}so, Chile}

%
%
\collaboration{The NA64 Collaboration}\noaffiliation
\vskip 0.25cm